\documentclass[usenatbib,usegraphicx]{mn2e}
\usepackage{epstopdf}
\usepackage{graphicx}
\usepackage{natbib}
\usepackage{amssymb}

\newcommand{\Lcr}{L$_{crit}$\ }
\newcommand{\Lcrns}{L$_{crit}$}
\newcommand{\Msol}{M$_{\odot}$}
\newcommand{\beq}	{\begin{equation}}
\newcommand{\eeq}	{\end{equation}}
\newcommand{\avg}[1]  {{\langle #1 \rangle}}

\defcitealias{kirk11}{KM11}
\defcitealias{gutermuth09}{G09}

\font\tenbi=cmmib10 
\newfam\bifam  \textfont\bifam=\tenbi

\font\tenbr=cmbx10
\newfam\brfam  \textfont\brfam=\tenbr

\font\squinttenbi=cmbx10 at 9pt
\scriptfont\brfam=\squinttenbi

\def\msun{\ifmmode {\rm M}_{\mathord\odot}\else $M_{\mathord\odot}$\fi}

\begin{document}

\title{The Formation and Evolution of Small Star Clusters}

\author[Kirk et al.]
{Helen Kirk,$^{1,2}$\thanks{Banting Fellow}\thanks{presently at National
	Research Council of Canada, Radio Astronomy Program} 
Stella S.~R.~Offner,$^{1,3}$\thanks{Hubble Fellow}
Kayla Redmond$^4$ \\
$^1$Harvard-Smithsonian Center for Astrophysics,
Cambridge, MA, 20138, USA\\
$^2$ Origins Institute, McMaster University, Hamilton, ON, L8S 4M1, Canada;
kirkh@mcmaster.ca\\
$^3$Yale University, New Haven, CT, 06511, USA; stella.offner@yale.edu\\
$^4$University of North Carolina at Asheville, NC  }
  
%\author{Helen Kirk\altaffilmark{1,2}, 
%Stella S.~R.~Offner\altaffilmark{1,3,4},
%Kayla Redmond\altaffilmark{5}}
%\altaffiltext{1}{Harvard-Smithsonian Center for Astrophysics,
%Cambridge, MA, 20138, USA}
%\altaffiltext{2}{currently a Banting Fellow at the Origins
%Institute, McMaster University, Hamilton, ON, L8S 4M1, Canada;
%kirkh@mcmaster.ca}
%\altaffiltext{3}{New Haven, CT, 06511, USA; stella.offner@yale.edu}
%\altaffiltext{4}{Hubble Fellow} %SSRO Must be specified as a footnote following HF grant requirements.
%\altaffiltext{5}{University of North Carolina at Asheville, NC}
%\author{Helen Kirk}
%\affil{Harvard-Smithsonian Center for Astrophysics,
%    Cambridge, MA 02138}
%\email{hkirk@cfa.harvard.edu }
%
%\author{Stella S. R. Offner}
%\affil{Harvard-Smithsonian Center for Astrophysics,
%    Cambridge, MA 02138}
%\email{soffner@cfa.harvard.edu }
%
%\author{Kayla Redmond}
%\affil{University of North Carolina at Asheville, NC}

\maketitle

\begin{abstract}
Recent observations show that small, young, stellar groupings
of $\sim10$ to 40 members tend of have a centrally-located most massive 
member, reminiscent of mass segregation seen in large
clustered systems.  Here, we analyze hydrodynamic simulations which form small
clusters and analyze their properties in a manner identical to the
observations.  We find that the simulated clusters possess similar
properties to the observed clusters, including a tendency to exhibit 
mass segregation.  In the simulations, the central location of the 
most massive member is not due to dynamical evolution, since there is little
interaction between the cluster members.  Instead, the most massive 
cluster member appears to form at the center. We also find that the 
more massive stars in the cluster form at slightly earlier times.
 
\end{abstract}

\section{Introduction}

Stars typically do not form in isolation but rather within groups of
a hundred or more stars \citep{lada03}. Even star-forming regions that have low-stellar
densities, e.g., Taurus, contain clear groups of 10 or more
stars (Kirk \& Myers 2011, hereafter KM11). 
The pervasiveness of stellar clustering from  
the very earliest embedded phase of star formation
(e.g., Gutermuth et al 2009, hereafter G09) 
to old, dense globular clusters (e.g., \citealt{meylan97})
indicates that the nature of stellar distributions holds clues
about the formation and dynamical evolution of stars.

In local star-forming regions, \citet{bressert10} find that the
distribution of protostellar separations has no characteristic scale.
One interpretation of this result is that there is no obvious preferred scale to separate
``clustered'' stars from ``non-clustered'' stars.  Simulations suggest that clustering can be dynamically erased rapidly or may, when present, provide no distinctive signature in the separation distribution \citep{gieles12,parker12}.
However, singling
out higher density concentrations is nonetheless instructive because   
the most densely distributed stars are most likely to gravitationally
interact and evolve as an ensemble.

A variety of definitions exist to dictate what is a stellar cluster. \citet{lada03}
propose a minimum
of 35 ``physically related'' stars with total mass density $> 1.0
\msun$ pc$^{-3}$. This definition is derived by requiring that a  
grouping survive evaporation for 10$^8$ years.
To be applied in practice, the \citet{lada03} cluster definition 
and other similar approaches \citep[e.g.,][]{Jorgensen08} 
use stellar surface density thresholds to determine cluster membership.
A number of authors have recently utilized a different method, the 
minimal spanning tree (MST), to define stellar groups \citep{cartwright04,
gutermuth09, allison09b, maschberger10, maschberger11, kirk11, parker12}.  
An MST is essentially a structure which connects all points together
through their minimum distances, much like the sketch of a
constellation.  We adopt the MST method for identifying 
stellar groups for easier comparison with observational results.
Ultimately, all methods work to identify regions of relatively
high stellar density 
but are subject to some subjective decision about the location
of the cluster boundary.

Inspection of the highest stellar density regions indicates that the most 
massive stars tend to be located closer to cluster centers and in the
highest stellar density regions \citep{hillenbrand08, gouliermis04}. This
appears to be true also for clusters where the most massive star is 
only a few solar masses \citepalias{kirk11}. Since it is only possible to
assign stellar mass after an age of a couple million years, at which point most of the natal
cloud gas has been accreted or expelled, it is difficult to determine the
primordial distribution of masses.  

Numerical simulations thus provide
an important avenue for exploring early cluster properties.
Early work modeled the dynamical evolution of small clusters by 
beginning with a set of stellar seeds in a gas potential \citep{bonnell97,
bonnell01,delgado03}. In later work, simulations also followed the formation 
of individual stars from the gravitational collapse of dense gas cores 
(e.g., \citealt{klessen01,bate03,bonnell04,Offner08a,girichidis11}).
However, even with the aid of simulations it remains unclear whether observed mass segregation is primordial or
dynamical and depends partially on the assumed model and
initial conditions \citep[e.g.,][]{bonnell98}.
For example, N-body simulations of clumpy, subvirial (i.e., collapsing)
clusters find that mass segregation can occur within a global
dynamical time of the full system \citep{allison09,parker13}.  
In contrast, \citet{maschberger10} find that
marginally bound hydrodynamic simulations produce mass segregation
within 0.5~Myr of formation.  \citet{girichidis12b} find that the
steepness of the initial large-scale gas density profile from which
stars are formed also affects the final spatial distribution of stellar
masses.

In principle, the details of clustering and mass segregation
can be used to differentiate between theoretical models. In
practice, however,
it is uncertain whether observed early mass segregation favors the turbulent
core model \citep{krumholz07} or competitive accretion model 
\citep{bonnell01}. 
In the former scenario, massive stars form from 
high-column density gas, which is incidentally often centrally
located. In the latter scenario, stars forming
at the cloud center are deep within the gravitational well and
thus have the largest reservoir of available gas, which allows them to
grow to higher masses.

Most simulation analyses have focussed on systems representing 
larger-cluster formation, including the formation of massive stars and
high stellar density environments.  We focus here on the less-explored
regime of the formation of small ($N$=10-40), sparse (1-10's $pc^{-2}$) 
intermediate mass ($M < 4$~\Msol) stellar groups comparable
to those in \citetalias{kirk11}.
In this regime, stellar dynamics might be expected to play a smaller role,
although \citetalias{kirk11} do observe mass segregation at an early age.
We also investigate the
effect of global cloud properties, such as Mach number, temperature, and
turbulent driving scale, on cluster properties. 

Our work extends previous studies in several important ways. First, 
we quantitatively define subclusters within the simulation and then follow 
the evolution of these groups and their properties over several cluster 
dynamical times for a range of cloud properties. Second, we apply 
observational constraints  and  quantitatively compare the simulated 
cluster properties with observations of similar clusters. Finally, we 
analyze simulations with continuous turbulent energy injection rather 
than simulations of isolated clouds wherein turbulence is allowed to 
decay. The former picture is more similar to the ``distributed" star 
formation in nearly regions, which have relatively low stellar surface 
densities and are not strongly centrally condensed.

In Section \ref{numsims} we describe our simulation parameter
study. In Section \ref{sec_mst} we discuss the identification of stellar
groups using minimal spanning trees (MST). We analyze the cluster
properties in Section \ref{results}, 
including the details of the mass distributions and member separations.
In Section \ref{sec_comp_obs} we compare with observations of young 
clusters. Finally, we summarize our conclusions
in Section \ref{conclusions}. 

\section{Numerical Simulations}\label{numsims}

We analyze six simulations of molecular clouds forming stars. We
perform all 
simulations with the ORION adaptive mesh refinement
(AMR) code \citep{truelove98, klein99}. The simulations do not include
magnetic fields and five assume a simple isothermal equation
of state. The simulation procedure is described in \citet{Offner09} 
and \citet{Offner13}, so we give only a brief summary here.

We initialize the simulations with uniform density and then
perturb the gas for two to three crossing times using a random velocity
field. This field has a flat power spectrum over wavenumbers $k_1$ to $k_2$
(see Table \ref{tab_simprops}), and we renormalize the perturbations to 
maintain a constant cloud velocity dispersion.  After 
the initial driving phase that produces a well-mixed turbulent distribution, 
we turn on gravity and allow collapse to proceed. 

The simulations each have a 256$^3$ base grid and four levels of AMR
refinement, where we automatically add new grids to satisfy the Jeans
criterion and adopt a Jeans number of 0.125 \citep{truelove97}. We introduce
a sink particle when the Jeans condition is violated on the finest
level \citep{krumholz04}. These particles approximately represent
young stellar objects, and henceforth, we shall refer to them as 
``stars.''  Since we do not include mass loss due to protostellar outflows, the simulated particle masses 
should be considered upper
limits. We also merge particles if they approach within four fine cells and if one has $ m_* \le
0.1~\msun$. The sink particle-gas interaction is smoothed on scales of one fine cell, such that the dynamics of closely approaching particles, especially when their mass is comparable to the cell gas mass, is not well modeled. In any event, we expect the formation of small fragments to be suppressed with the addition of either radiative transfer or magnetic fields 
\citep[e.g,][]{Offner09,commercon11}. Thus, our 
stars may also sometimes represent binary or multiple 
star systems rather than individual stars.

Table \ref{tab_simprops} displays the simulation parameters. The parameters
of the fiducial Rm6 simulation are set such that the cloud is virialized:
\begin{equation}
\alpha = \frac{5 \sigma^2 (L/2)}{GM} = 1,
\end{equation}
where $L$ is the cloud size, $M$ is the cloud mass, and $\sigma = \mathcal{M} c_s/\sqrt{3}$ is the one-dimensional velocity dispersion, 
where $c_s$ is the sound speed.
The fiducial simulation also obeys the observed linewidth-size relation
\citep[e.g.,][]{solomon87,mckee07}: 
\begin{equation}
\sigma = 0.72 \left( \frac{R}{1{\rm pc}}\right)^{0.5} {\rm km~s}^{-1}. 
\end{equation}
The other simulations vary in temperature, Mach number,
driving scale, and physics. Table \ref{tab_simprops} illustrates that the
differences lead to significant statistical differences in the
number density of stars. For reference, simulation Rm6s has parameters
identical to Rm6 but uses a different turbulent random seed and produces 
a similar surface density of stars, $n_*$, to the Rm6 simulation.

\begin{table*}
\begin{minipage}{126mm}
\caption{Simulation Properties \label{tab_simprops}}
\begin{tabular}{llllllllll}
%HK added final columns as per Thomas Maschberger's suggestions
%SSRO supplied values for very last 2 columns
\hline
Run$^a$ & 
$L$$^a$  & 
$M$$^a$ &
T$^a$ &
$\mathcal{M}$ $^a$ &
$k_1..k_2$$^a$ & 
$n_*$$^b$ &
t$_{end}^c$ &
$N_{*}^c$ & $SFE^c$ \\
 & (pc) & ($\msun$) & (K) & & & (pc$^{-2}$) & (Myr) & & (\%) \\
\hline
Rm6$^d$ &2 & 600 & 10 & 6.6 & 1..2 & 18.2 & 0.95 & 88 & 17.6\\ %73 %18.4 74
Rm6s &2 & 600 & 10 & 6.6 & 1..2 &  18.8 & 0.95 & 100 & 18.1\\%75  %19.0 76
Rm9$^d$ &2 & 600 & 10 & 8.9 & 1..2 &  3.2 & 0.95 & 17 & 3.5\\ %same 13
Rm4$^d$ &2 & 600 & 10 & 4.2 & 1..2 & 13.8 & 0.89 & 80 & 10.4\\ %55 %14.0 56
Rk34 &2 & 600 & 10 & 6.6 & 3..4 & 3.2 & 0.91 & 14 & 2.6 \\%13  %3.6 14
Rt20 &2 & 600 & 20 &  6.6 & 1..2 &  1.5 & 0.95 & 17 & 3.8\\%6 %4.4 17
Rrt$^e$ &0.65 & 185 & 10 &  6 & 1..2 &  30.6 & 0.32 & 18 & 7.2\\ % same 13
\hline
\end{tabular}
$^a$Simulation ID, box length, total initial gas mass,
  initial gas temperature, and Mach number, respectively.\\
$^b$The projected number density of stars after one
  freefall time including only those stars with $m_*> 0.03 ~\msun$. \\
$^c$The time elapsed in the simulation after gravity was turned on,
  and the total number of stars, and star formation efficiency (stellar
  mass divided by total mass) at this final time. \\
%Note to HK: these densities are not scaled by 1/3
$^d$Properties of this simulation were previously analysed in \citet{Offner13}.\\
$^e$This simulation from \citet{Offner09} includes radiative transfer; it
  models heating due to accretion and nuclear burning.
\end{minipage}
\end{table*}

\section{Cluster Identification}\label{sec_mst}

\subsection{Simulated Stellar Catalogs}

Our method of cluster identification was designed to mimic that in
\citetalias{kirk11}.  The location of stars formed within the simulations
was first projected onto three planes ($xy$, $xz$, and $yz$), as an observer
would view the simulated region from a large distance away along the $z$, $y$, 
and $x$ axes.  Since the simulations are performed within a periodic box,
we replicate the simulated stellar distribution to a total of 
three by three boxes,
to ensure any clusters with members across a box edge are
properly identified.  In our final analysis, only clusters 
identified which have centres within the inner box are included,
to ignore duplicate clusters.

In real molecular clouds, dense, star-forming cores are estimated
to lose about two-thirds of their mass to protostellar outflows and
winds before the final mass of the star is set,
assuming that there is a one-to-one mapping between the 
dense core and initial stellar mass function \citep[e.g.,][]{Alves07}.
Since outflows are not included in these simulations,
we reduce the masses of the sink particles formed by two thirds;  
the exact factor used has little effect on our final results, 
the majority of which involve relative stellar masses.
We eliminate any sink particles which have re-scaled masses
of $< 0.03$~\Msol.
Observationally, YSOs at very low masses become
difficult to detect; the surveys analyzed by \citetalias{kirk11} suffer
from incompleteness around 0.01 to 0.03~\Msol.  
Finally, in each of the three projected views, we remove
the lower mass member of any pairs separated by less than 1000~AU.
This separation corresponds to the approximate resolution limit of
the stellar catalogs analyzed by \citetalias{kirk11}.  Since the observed
stellar masses were derived from spectral types, any
low-mass companions closer than the minimum resolution would
not have been counted in the observational mass estimate.
Note that our results are not strongly sensitive to the precise choice
in cutoff values; the main effect is to slightly change the value of
\Lcr used to define clusters (discussed in the following section).
In Appendix~A, we demonstrate that our final results are robust to
changes in \Lcrns.
We note that the simulations without radiative feedback 
likely over-estimate the amount of fragmentation on small scales 
\citep{Offner09,Bate09}.  By excluding very small stars and the smaller member 
of close pairs, which we do in order to adhere to observational limits, 
we also correct for this effect.

\subsection{MST Identification}

\begin{figure*}
\begin{tabular}{ccc}
\includegraphics[width=2.2in]{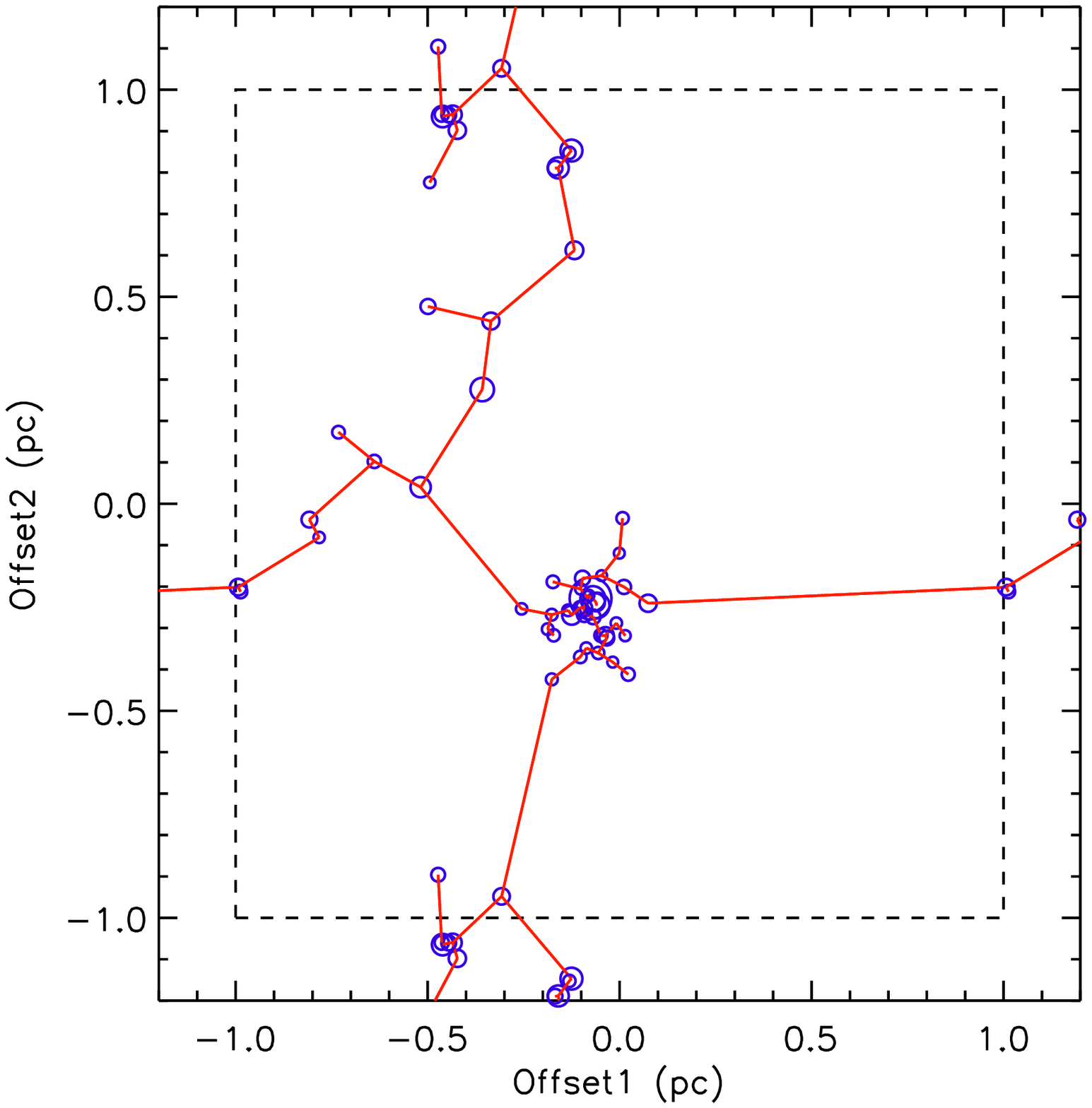} &
\includegraphics[width=2.2in]{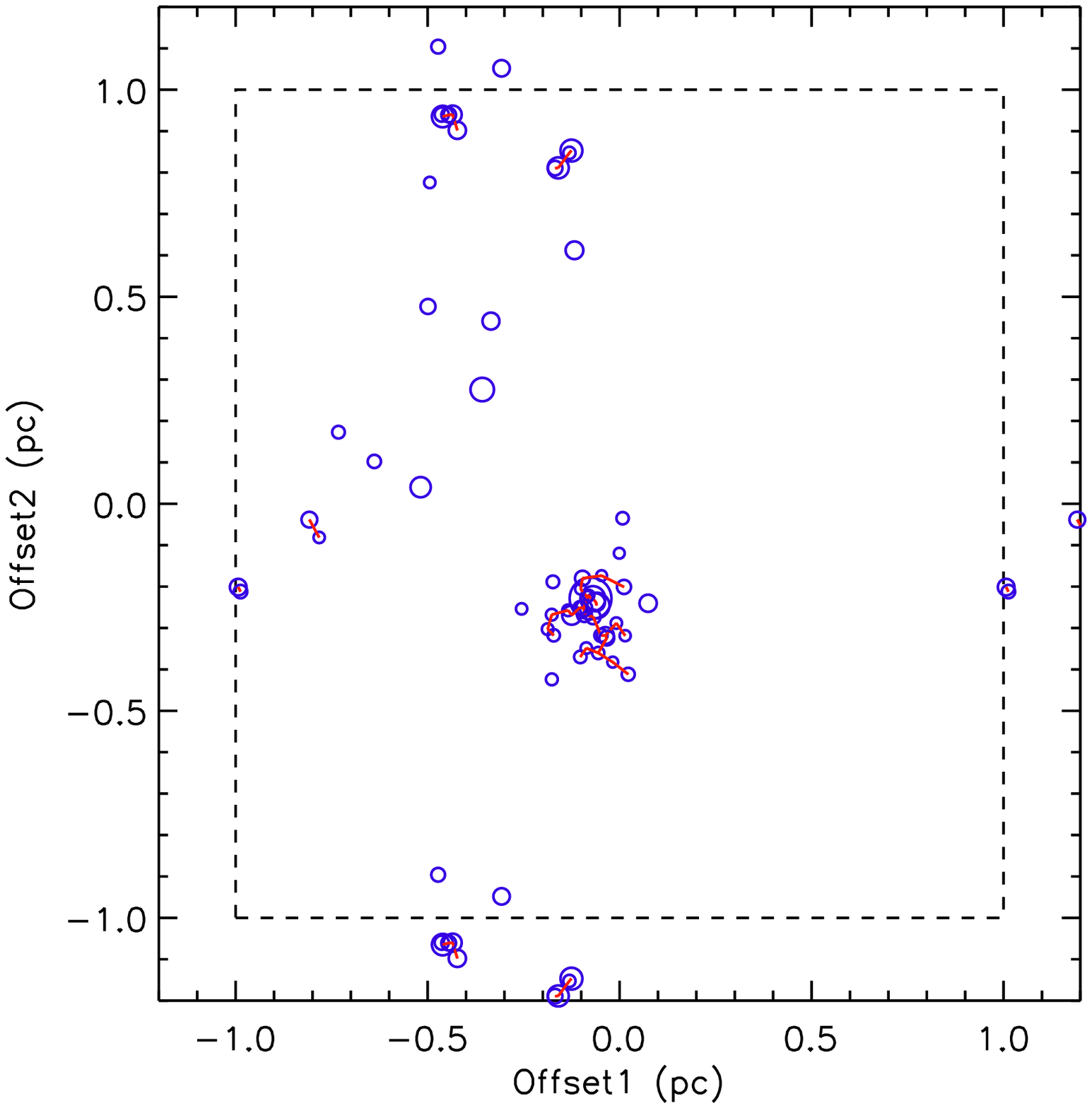} &
\includegraphics[width=2.2in]{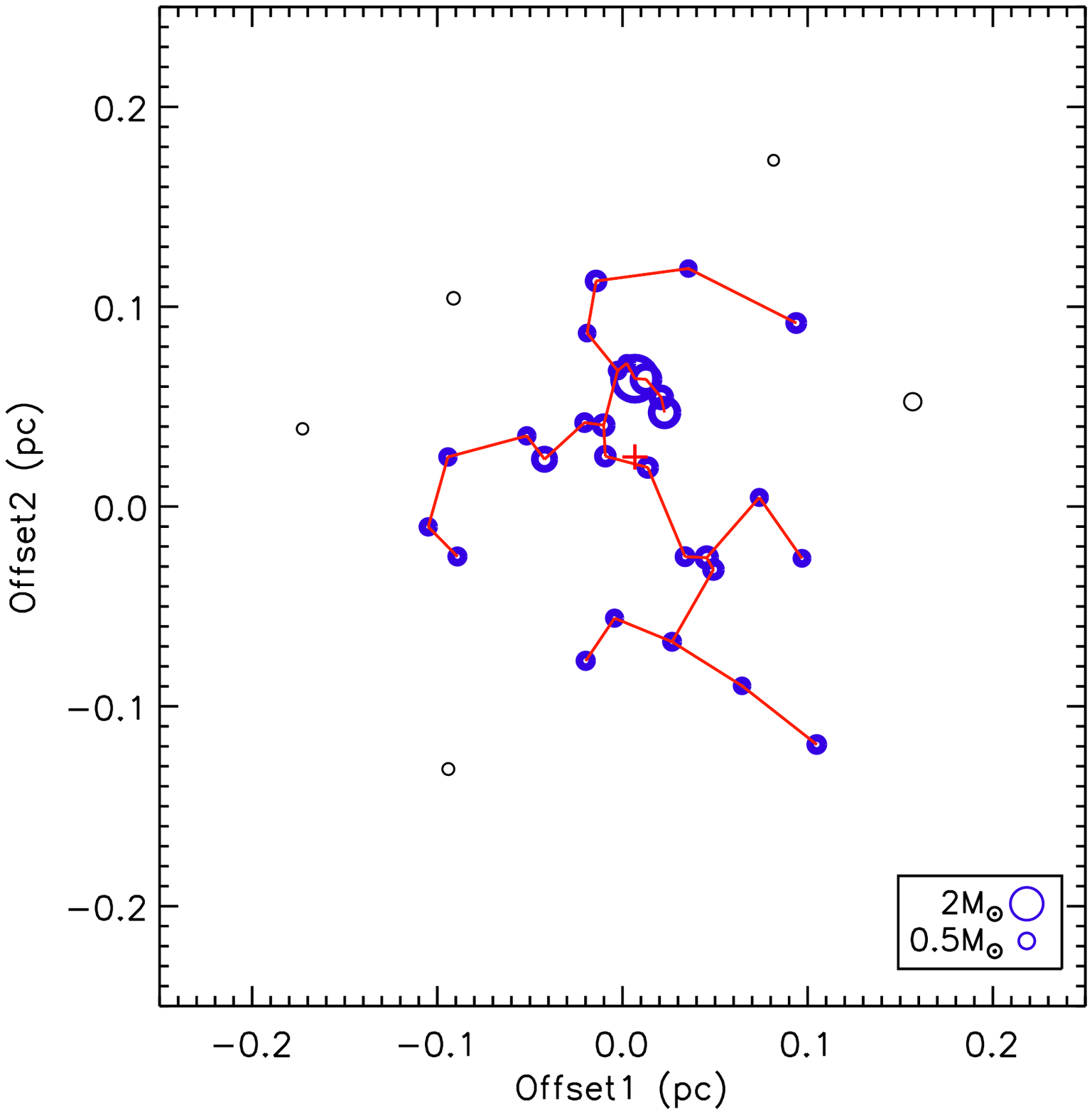} \\
\end{tabular}
\caption{The minimal spanning tree structure computed for the Rm6 simulation 
	at the final time step (0.95~Myr) when viewed along the $z$ axis ($xy$ 
	projection).  The original simulation is contained within the dashed 
	square, while the outer borders show part of the replicated boxes.
	The blue circles show the locations of the sink particles, with their 
	size scaling with mass (see legend in right panel), 
	while the red lines show the MST structure.
	Note that 
	additional clustered structure occurs across the original
	simulation boundaries.  Left: The full MST structure. Middle: The 
	MST structure remaining after branches with lengths larger than 
	$L_{crit}$ (here 0.07~pc) have been removed.  Groupings of
	more than ten sink particles which remain connected are
	classified as clusters.  Right: A close-up view
	of the single cluster found in this example.  Here, non-members
	are shown in black, and the cluster centre is indicated by
	the red plus sign.
	}
\label{fig_mst_full}
\end{figure*}

With this final set of simulated stellar catalogs, we identify
small stellar clusters for each of the three projections using
minimum spanning trees.
An MST represents
the minimum total length by which all points can be connected
\citep[e.g.][]{Barrow85}; each connecting line segment is referred to 
as a branch.  
Once the full MST structure is calculated, we follow
the procedure of \citetalias{gutermuth09} and \citetalias{kirk11}
to identify clusters.  Figure~\ref{fig_mst_full} shows an example
of this full procedure.

In the left panel of Figure~\ref{fig_mst_full}, the full MST structure
is shown for the $xy$ projection of the Rm6 simulation at its final
time step (0.95~Myr).  
Since the simulation has been replicated three times in either direction to
allow us to account for the effect of periodic boundaries,
several structures can be seen more than once.  For
clarity, only a small portion of the outer replicated 
simulation boxes are displayed.

Within the full MST structure, clusters are visually apparent as 
regions of small star-star separations, i.e., 
stars connected by short branches.  Clusters can therefore
be defined as stars all interconnected by branch lengths less
than some value, the critical branch length, \Lcrns.
\citetalias{gutermuth09} found that nearby clustered regions
have MSTs with a characteristic cumulative distribution of 
branch lengths: a steep rise at small branch lengths followed
by a turn-over to a shallow slope, illustrated in
Figure~\ref{fig_brlens_Rm6seed}.
Given this characteristic cumulative branch length distribution,
the branch length of the turn-over is an obvious choice for
\Lcrns, and that is effectively what \citetalias{gutermuth09}
and \citetalias{kirk11} 
 adopt.  Appendix~\ref{app_critlen} discusses
the determination of \Lcr in more detail.  With such
a procedure, regions which have different mean clustering properties
have different \Lcr values, which more naturally captures the  
local stellar overdensities than adopting a constant value would
do.  In \citetalias{kirk11}, there is a factor of a few difference between
\Lcr found for the sparse stellar distribution in Taurus compared with
the much denser clustered IC~348.

For our analysis of the simulations, if we followed an identical
method, we would potentially have a different \Lcr measured
for each simulation, at each time step, in each of the three 
viewing angles.  With the analysis in Appendix~\ref{app_critlen},
we demonstrate that it is sufficient for our purposes to adopt
a single constant value of \Lcr for each simulation (i.e.,
no variation with time or viewing angle).  We do not find strong
evidence for real variations in \Lcr within each simulation,
and maintaining a single \Lcr allows for easier intercomparisons
within each simulation.  Allowing for a different \Lcr between
the various simulations is important, as different initial conditions
can result in different clustering properties.  The \Lcr adopted
for each simulation is given in Table~\ref{tab_critlens}.

\begin{table}
\caption{MST statistics \label{tab_critlens} }
\begin{tabular}{lcccccc}

\hline
Sim &
L$_{crit}$ $^a$ &
Median$^b$&
Mean$^b$&
Min$^b$&
Max$^b$&
L$_{rat}$$^c$ \\
 &
(pc)&
(pc)&
(pc)&
(pc)&
(pc)&
 \\
\hline
%Lcrits are from the var_MST results.  The rest are a result of averaging
% (or taking the minimum / maximum) of the values for the 3 projections
% at the final time step for each simulation, in 
% fixed_MST/[sim]/[time]_[critlen]_MST_stats.txt
%UPDATE: Final column (Lcr/sep_max) is from programs/measure_Lratio.txt
Rk34 & 0.12 & 0.21  & 0.34 & 0.007 & 1.2 & 0.027 \\
Rm4  & 0.06 & 0.05 & 0.13 & 0.004 & 1.1 & 0.013 \\
Rm6  & 0.07 & 0.04 & 0.10 & 0.005 & 0.95 & 0.016 \\
Rm6s & 0.07 & 0.05 & 0.09 & 0.005 & 0.6 & 0.013 \\
Rm9  & 0.1$^d$ & 0.40 & 0.37 & 0.005 & 0.75 & 0.022 \\
Rrt  & 0.1$^d$& 0.12 & 0.12 & 0.016 & 0.3 & 0.061 \\
Rt20 & 0.1 & 0.14  & 0.25 & 0.005 & 1.2 & 0.025 \\
\hline
\end{tabular}
$^a$The critical branch length used for identifying clusters
	in the simulation (all projections, all time steps).\\
$^b$The median, mean, minimum, and maximum branch lengths
	in the final time step of each simulation, averaged over all
	three projections.\\
$^c$The ratio of $L_{crit}$ to the maximum projected separation
	between sources in the final time step of each simulation,
	averaged over all three projections.\\ 
$^d$$L_{crit}$ is poorly determined due to the small number of
	sources.
\end{table}

As can be seen in Table~\ref{tab_critlens}, \Lcr varies by a factor
of about two between the different simulations, ranging from
0.06~pc to 0.12~pc.
In \citetalias{kirk11}, the relatively sparse Taurus
stars have an \Lcr of 0.52~pc, while the
tightly clustered stars in IC348 have an \Lcr
of 0.083~pc; \citetalias{gutermuth09} found \Lcr values
between 0.086~pc and 0.7~pc in their sample.
The values of \Lcr measured in the simulations span a smaller
range than these two observational surveys, and tend to be 
small.
The \Lcr values are, however,
larger than that used in \citet{maschberger10} (of 0.025~pc)
for the analysis of a competitive accretion simulations showing
the formation of a larger, more densely clustered region.
The MST analysis in \citet{girichidis12b} is restricted
to the $\Lambda_{MST}$ method of \citet{allison09b}, which,
in the case of large-$N$ clusters without substructure,
can be an effective method to measure mass segregation.
The $\Lambda_{MST}$ method 
does not require \Lcr to be determined, however, a small \Lcr would be
expected, given the high stellar
densities in the simulations.

A second statistic associated with \Lcr is the ratio of \Lcr
to the maximum separation between stars.  In observed clusters,
regions with sparser clusters such as Taurus tend to have a larger \Lcr
values as well as a larger maximum separation between stars.  
In denser clusters, both \Lcr and the maximum separation between stars
are smaller, and \Lcr (and by definition, also the maximum separation)
has also been found to increase when a more extended, sparser,
region around a cluster is included in the analysis \citep{Masiunas12}.
We find that the ratio of \Lcr to the maximum separation between stars is
relatively constant in the observations, ranging from 1\% to 3\% for the
clusters in \citetalias{kirk11}.  Performing a comparable measurement on the
simulations reveals that they are generally consistent with the
observations, as shown in
Table~\ref{tab_critlens}.  All of the simulations except for Rrt
have ratios between 1\% and 3\%, while in Rrt, the ratio is 6\%.
Rrt was one of two simulations where the value of \Lcr measured was
highly uncertain, so the poorer agreement with observations is perhaps
unsurprising.

Once \Lcr is determined, and the initial full MST structure is 
``pruned'' of all branches longer than
\Lcrns (middle panel of 
Figure~\ref{fig_mst_full}); stars which remain connected through
short branches are potential clusters.  \citetalias{kirk11} impose
an additional criterion that clusters have more than ten 
members to allow for meaningful analysis of cluster properties.
The right panel of Figure~\ref{fig_mst_full} shows a zoomed-in
view of the cluster identified in the middle panel with
the red plus showing the cluster's centre.
As in \citetalias{kirk11}, we 
define the centre as the median position of all
cluster members; this is 
less prone to bias than, e.g., the mass-weighted mean position
for measurements of mass segregation.  In Appendix~\ref{app_meancentre}
we demonstrate that using instead the mean cluster member position as the
cluster's centre has little impact on our results.

\section{Results}\label{results}

\subsection{Mass Distribution}
\begin{figure}
\includegraphics[width=3.5in]{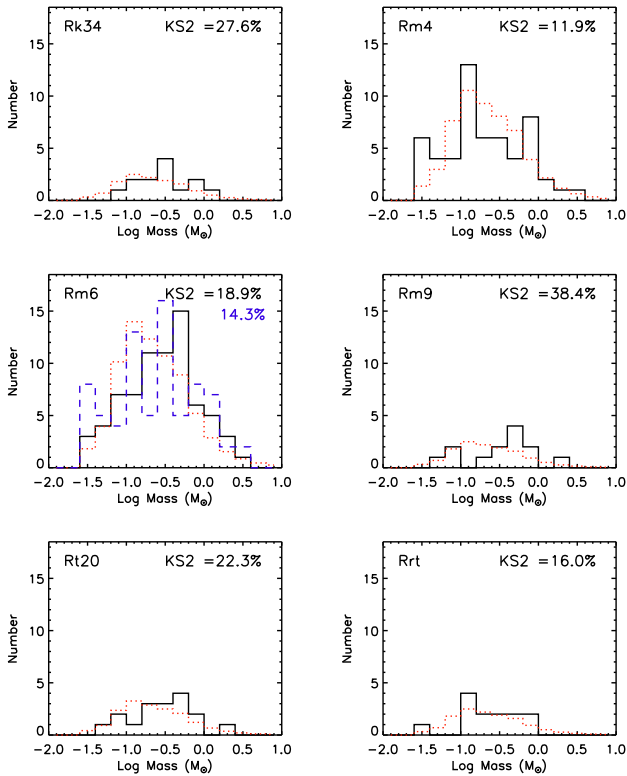}
\caption{The distribution of stellar masses at the final time
        step of each simulation.  The two runs of Rm6 with differing
        initial turbulent seeds are plotted in the same panel,
        with Rm6s shown by the dashed blue line.  The dotted red
        line in each figures shows the the Kroupa IMF distribution
	\citep[as given in][]{Weidner10},
        scaled to the same total number of stars, and given the same
        minimum mass cutoff.  The numbers in the top right corner
        show the probability that the mass distribution in the
        simulation is drawn from the same parent sample as the
        Kroupa IMF.}
\label{fig_massdistr}
\end{figure}

Figure~\ref{fig_massdistr} shows the stellar masses
at the final time step in each simulation, with the masses reduced by
a factor of 3 from the sink cell masses, and the observational limit
of 0.03~\Msol\ applied.  Earlier time steps
generally follow the same trend but with fewer sources.
Observed young stellar mass functions tend to be consistent with the standard
IMF, with the possible exception of Taurus, which has a relative excess 
of sub-solar, K7-M1 type stars \citep[e.g.,][]{Luhman09}.
Figure~\ref{fig_massdistr} shows the stellar mass
distributions (solid black lines) compared to the Kroupa IMF
(red dotted line).  
The simulation mass distributions are consistent with a Kroupa
IMF; a two-sided Kolmogorov-Smirnov test \citep[e.g.,][]{Conover99}
gives probabilities between 10 and 40\% that the two are drawn from
the same parent sample.  We note that earlier time steps in the 
simulations give a similar result, and the two runs of Rm6 are
statistically similar.  For reference, the Jeans mass of Rm6 is $M_J=8.0 \msun$, where $M_J = 4/3\pi \bar \rho (L_J/2.0)^3$ and the Jeans length is defined as $L_J= c_s[\pi/(G \bar \rho)]^{1/2}$. This is slightly higher than the mean stellar mass we find in the simulations.

\subsection{Mass Evolution}
\begin{figure}
\includegraphics[width=3.5in]{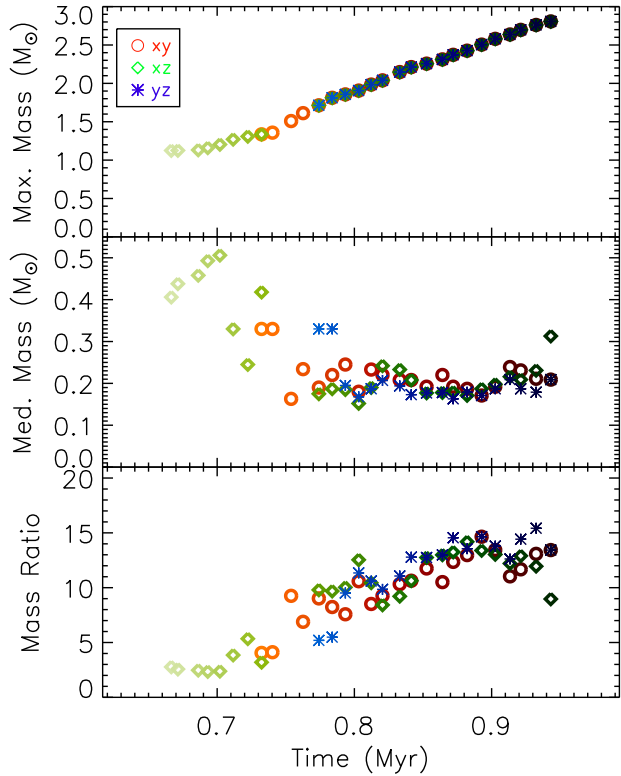}
\caption{The variation in maximum mass, median mass, and mass
	ratio in the Rm6 simulation as a function of time.
	Red circles, green diamonds, and blue asterisks show
	the results for the three projections, and earlier times
	are shown with lighter colours.  The simulations were 
	sampled every 0.01~Myr to provide more readable plots.}
\label{fig_massratio_time}
\end{figure}

We next examine the stellar mass properties within each cluster
identified.  Figure~\ref{fig_massratio_time} shows the evolution
of the mass of the most massive cluster member 
for the Rm6 simulation, for each of the three projections, every 0.01~Myr 
(top panel).
Note that only one cluster with more than 10 members was identified in
each projection.
The most massive star formed in this simulation is located within a
bona-fide 3D clustering of sources, and hence the same point is plotted
for the clusters identified in all three projected views.  At early times,
the cluster has fewer members, and due to the small motions of the stars
near the cluster's periphery, the grouping
tends to alternate between meeting and not meeting the cluster definition 
(more than 10 members all connected by MST branch lengths of less than 
\Lcrns).  This leads to some apparent gaps in the time-sampling plotted.

Figure~\ref{fig_massratio_time} shows that the most massive cluster 
member tends to grow with time due to accretion, in this case, following 
a roughly linear trend.  
Accretion also occurs in the other stars in the simulation, however,
this is less apparent when examining the median cluster member
mass as a function of time (middle panel of Figure~\ref{fig_massratio_time}).
At earlier times in particular, this accretion signature is masked by the
addition of new cluster members.  The majority of new cluster members
are new stars which formed in the simulation (or stars which have now
accreted sufficient material to surpass the observational threshold),
and their addition to the cluster tends to decrease the median cluster mass.
This effect is largest at the earlier times when the total number of
cluster members is small.  Other simulations have also been shown to 
produce time-independent IMFs that agree with the observed IMF \citep{krumholz12,bate12}. This is generally 
attributed to either stellar feedback or dynamical interactions regulating 
stellar masses. Since the IMF is observed to be robust to variation across  
a broad range of initial conditions and environments \citep{bastian10}, it 
is unlikely to have a preferred characteristic timescale.

The ratio of the mass of the
most massive and median cluster members (the ``mass ratio'' shown 
in the bottom panel of Figure~\ref{fig_massratio_time}) gives
a rough sense of how gravitationally dominant the most massive
cluster member is expected to be in terms of the cluster's dynamical
evolution.  
We find that the mass ratio tends to increase slowly with time,
spanning a slightly smaller range of values than the mass ratios
found in \citetalias{kirk11}.  This is discussed in more detail in  
Section~\ref{sec_comp_obs}.

\subsection{Separation Distribution}
\begin{figure}
\includegraphics[width=3.5in]{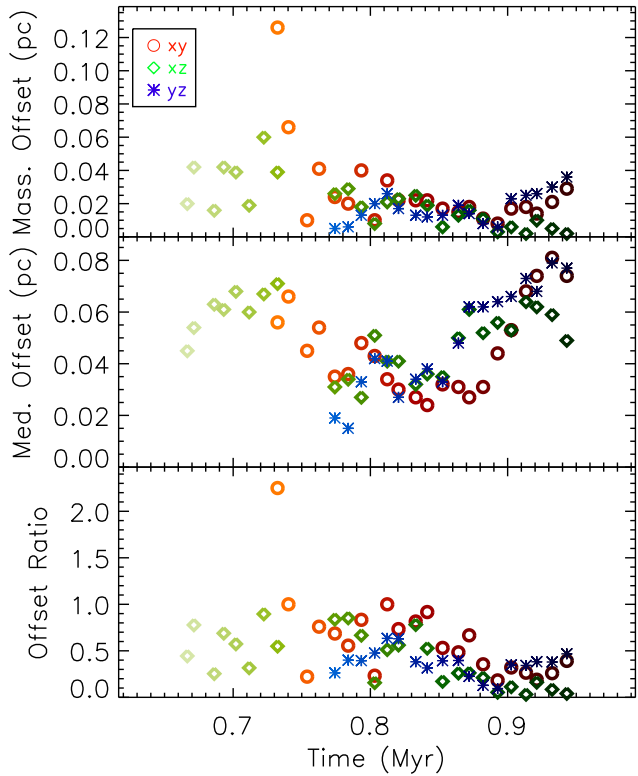}
\caption{The variation in offset from the cluster centre for
	the most massive member, the median cluster value,
	and the ratio of the two (the `offset ratio') in the Rm6 simulation
	as a function of time, sampled every 0.01~Myr.
	Red circles, green diamonds, and blue asterisks show the
	clusters identified in each of the three projections,
	and lighter shades indicate earlier time steps. 
	}
\label{fig_sepratio_time}
\end{figure}

We similarly analyze the evolution of the spatial positions
of the cluster members.  The top panel of 
Figure~\ref{fig_sepratio_time} shows
the evolution of the separation of the most massive cluster member
in the Rm6 simulation from their cluster centres.  From this figure,
it is clear that the separation neither monotonically increases nor
decreases -- rather, it is variable.  Much of this variability,
and all instances of large changes between adjacent time steps, 
are caused by variations in cluster membership, rather than substantial
motion of the most massive cluster member.
Particularly at earlier times, when the total number of cluster
members is small, the addition or subtraction of a cluster member
on the outskirts (with a separation of roughly \Lcr to its nearest
neighbour), can cause a significant change in the cluster centre position.
On rare occasions, a star near the
cluster boundary may itself be separated from one or two
additional non-cluster members by less than \Lcrns, in which
case, when it becomes less than \Lcr from an existing cluster member,
either through motion or the appearance of a new star inbetween,
the cluster gains several new members, and the cluster centre
is more strongly influenced.  The reverse situation can also
occur -- one or several cluster members near the outskirts which are 
separated by nearly \Lcr from the next cluster members may move slightly
farther away, reducing the overall cluster membership and
changing the cluster centre accordingly.
The motion of the most massive cluster member itself, contributes relatively
little to the changes observed in Figure~\ref{fig_sepratio_time},
and we discuss this in more detail in the following section. 

The middle panel of Figure~\ref{fig_sepratio_time} shows the 
variation in the median value of all cluster member's offsets
from the centre with time.  This variation is primarily
due to the change in the cluster centre, induced by cluster membership
changes. 
As with the most massive member's offset, the large,
sudden variation tend to be when multiple stars became members (non-members)
due to small motions taking them below (above) the \Lcr separation
from the next nearest cluster member.  Several of the large-scale
variations in the median offset can also be seen in the massive star
offset.

The bottom panel of Figure~\ref{fig_sepratio_time} shows the 
ratio of the most massive member's offset to the median offset,
the `offset ratio',
as a function of time.  This ratio gives an indication of how
centrally-located the most massive cluster member is: ratios much
less than one imply that the most massive cluster member lies near
the cluster centre, while ratios much larger than one would
imply the opposite; randomly located most massive members 
tend to have offset ratios around one \citepalias{kirk11}.  
The offset ratio shown
in Figure~\ref{fig_sepratio_time} shows significant scatter,
although slightly less than either the most massive star's or median
stars' offsets individually,
since both can be affected by changes to the 
location of the cluster centre in a similar manner.
It is notable that the offset ratio tends to be less than 
one at all times in the simulation, for all projections, and
there is no obvious evolution with time.
These points will be discussed in more detail in Section~\ref{sec_comp_obs}.

\subsection{Dynamical Evolution of Cluster Members}
\label{sec_dyn}
The previous results show significant mass segregation from the earliest 
times clusters are identified.  A cluster by our definition, however, must 
have a minimum of eleven members, so it is possible 
that some dynamical evolution occurs prior to the times analyzed above. To 
assess the amount of dynamical evolution, in this section we investigate the 
motion of the most massive cluster members beginning from their time of 
formation. If the most massive member moves significantly and  
interacts dynamically with other stars, then the mass segregation displayed 
by the MST clusters may not in fact be primordial.

There are two timescales that are relevant to the process of 
dynamical mass segregation. 
An estimate for the mass segregation timescale is given by the time it takes the most massive member to migrate to the cluster center. For a star of mass $M$ in a cluster with $N$ stars this can be expressed:
\beq
t_{\rm seg}(m) \simeq \frac{\avg{M}}{M}t_{\rm relax} \simeq \frac{\avg{M}}{M} \frac{N}{8 {~\rm ln} (N)} \frac{R}{\avg{v}},
\eeq
where $\avg{v}$ is the average velocity of a star and $R$ is the cluster radius.
\citep{spitzer69, allison09}.
For stars initially moving with $\avg{v} \simeq c_s$, $N=15$, 
$\avg{M}=0.6 \msun$\ and $R=0.2$ pc, a star of $3~\msun$\ will 
migrate to the cluster center in $t_{\rm seg} \sim 0.14$ Myr.  

This estimate implicitly assumes that the system is gas free and 
that the gravitational potential is dominated by the stars. However, the   
simulated systems here are gas dominated, which acts to damp the effect of two-body 
interactions. Overall, less than 20\% of the total gas in the simulations turns 
into stars by 1 $t_{\rm ff}$, although the volume restricted to the cluster
itself likely has a higher mass fraction in stars.
In a gas dominated system, the characteristic dynamical 
timescale can be approximated by the freefall time:
\begin{equation}
t_{\rm ff} = \left( \frac{3 \pi}{32 G \rho}\right)^{1/2},
\end{equation}

where $\rho =\mu_p n$ is the mass density and $\mu_p=2.33 m_{\rm H}$ is the mean particle  
mass. For a mean clump number density of $n =10^4$ cm$^{-3}$, 
$t_{\rm ff} = 0.34$ Myr, which is slightly longer than the N-body 
dynamical times. Protostars form from the dense gas 
($n>10^5$ cm$^{-3}$), which has a small relative offset velocity from 
the lower density core envelope gas (e.g., \citealt{kirk07, Offner09b,
kirk10}), so a single star shouldn't have a 
significantly different velocity than the gas it forms from. 
The core itself, however, may have some advection velocity, $t_{\rm adv}$, that 
is comparable to the sound speed or slightly higher. Protostars retaining 
their natal sonic velocity might migrate $\sim 0.2$ pc in 0.5~Myr.

These times are significantly shorter than the time over which the 
simulations form stars ($\sim$ 0.6-1.0~Myr), so in principle, the stars 
have sufficient time to become mass segregated due to dynamical 
interactions.  We examine this point in both the observed
projected view of clusters and the full 3D view.

\begin{figure*}
\begin{tabular}{c}
\includegraphics[width=7in]{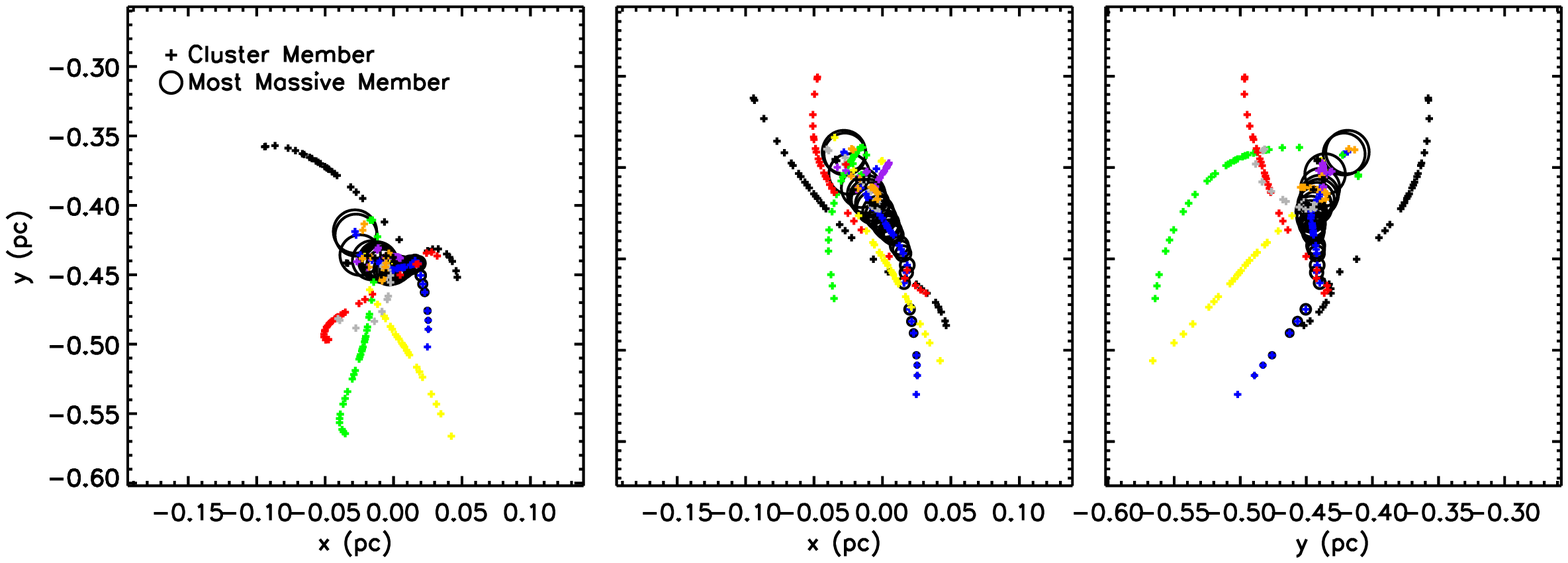} \\
\includegraphics[width=7in]{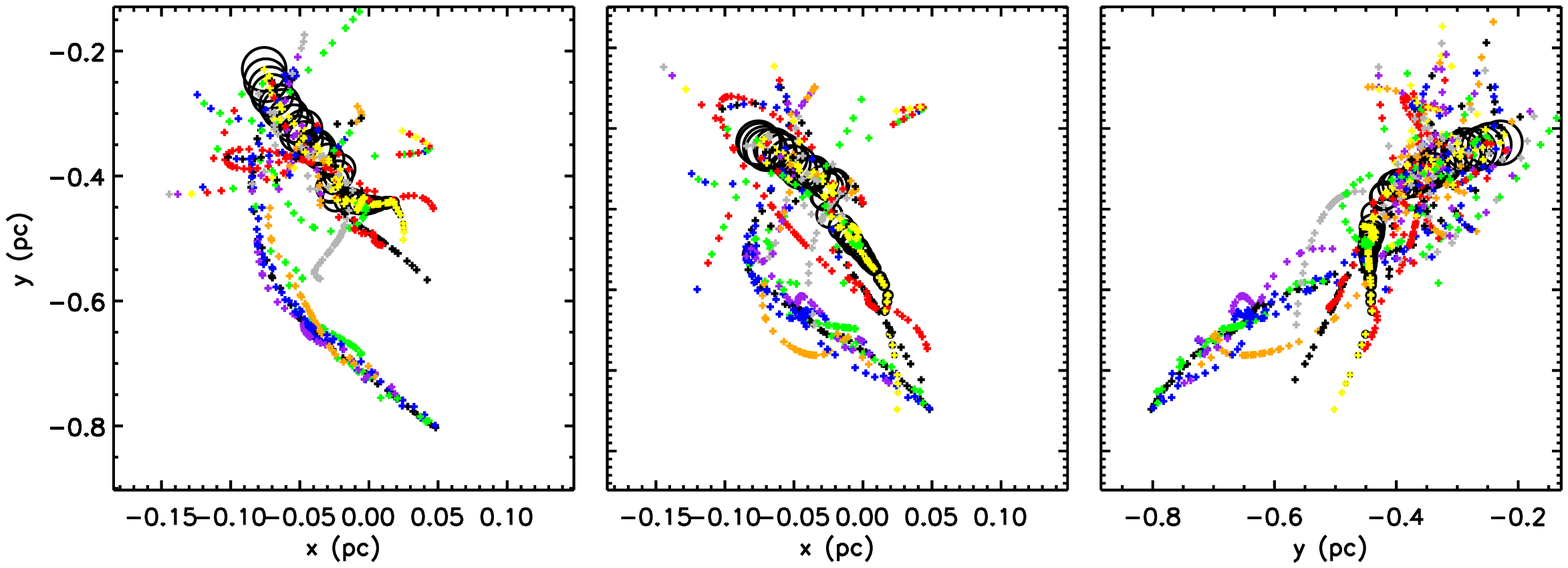} \\
\end{tabular}
\caption{The evolution of the projected member positions along the $x$, $y$, 
	and $z$ axes for each cluster member as a function of time for two 
	clusters in Rm6. The most-massive member is indicated with a black 
	circle that grows as a function of time. The top cluster was 
	identified from the $y$ projection (middle panel) and the bottom 
	cluster was identified from the $z$ projection (right panel).
	Positions are shown only every 0.01~Myr for clarity.
	}
\label{2dclusters} 
\end{figure*}

Figure~\ref{2dclusters} shows the projected positions of two clusters 
as a function of time for views along the $x$, $y$, and $z$ axes
($yz$, $xz$, and $xy$ projections). 
In these figures, we take the list of cluster members at 
a given time and projection and plot the 2D views of the cluster 
member positions sampled every 0.01~Myr.
Cluster member positions are indicated by the coloured plus
signs, while the most massive cluster member is shown with the
circle, whose size scales with time. 
The direction of motion of the most massive member can be 
inferred by following the 
circle from small to larger sizes. 
The top panel shows a small cluster which only satisfied all MST cluster
criteria for several time steps in the $xz$ projection.  
Several points
can be taken from this plot: first, the three projected views have
a fairly similar appearance, implying that many of the cluster members
are indeed clustered in 3D space.  Second, many of the stars remain
clustered at early times in the simulation -- the largest motion seen
is the bulk motion of the entire group.  The most massive
cluster member moves very little with respect to the surrounding
stars: only several lower mass stars have significant
relative motions as they join (or are expelled from) the cluster.

The lower panel of Figure~\ref{2dclusters} shows a similar view for
a larger cluster, identified in the $xz$ projection, which satisfied the
MST cluster definition for a large fraction of the simulation in multiple views.
These views are `messier', showing greater interactions
between cluster members.  Despite this, the most
massive cluster member usually appears near the middle of the 
stars, rather than moving there due to early dynamical interactions.
Finally, note that this figure illustrates that (larger) clusters may
include members which are only
associated along the line of sight: in the bottom left panel of 
Figure~\ref{2dclusters}, at earlier times (lower right of plot), there
are clearly two highly separated groupings of stars (similar position
in $x$ but not $y$), which appear as a single grouping in the middle plot
(similar $x$ and $z$ at the lower right).
Figure~\ref{2dclusters} also demonstrates that the cluster evolution
proceeds without the merging of small clusters \citep[e.g.,][]{maschberger10}, although individual stars may join. 
The stellar densities in the simulations are sufficiently low that only
1 or 2 clusters are present at once, so it is unlikely for clusters
to interact and merge.

\begin{figure}
\begin{tabular}{c}
\includegraphics[width=3.5in]{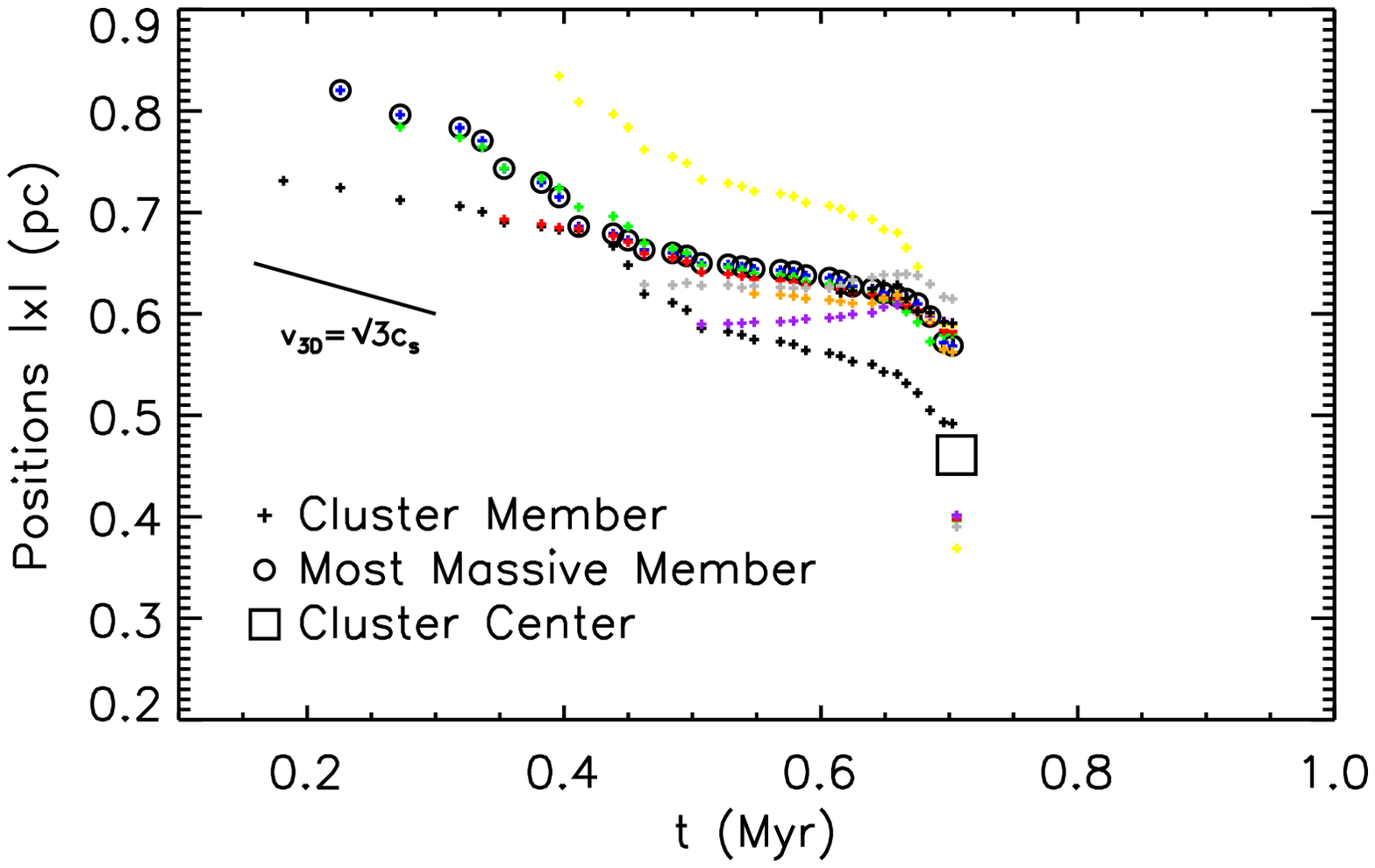}\\
\includegraphics[width=3.5in]{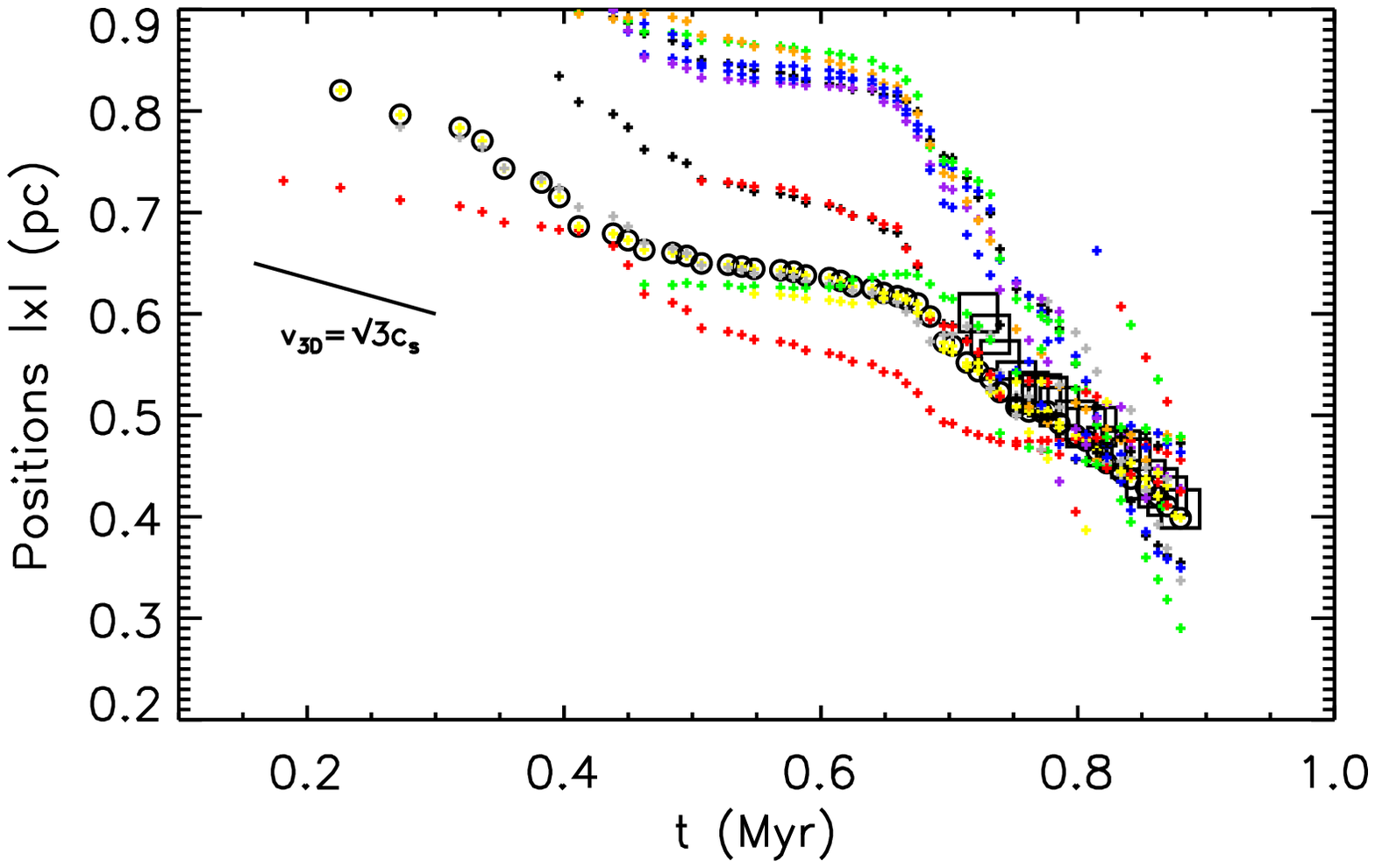}\\
\end{tabular}
\caption{The three-dimensional position ($|{\bf x}|=\sqrt{x^2+y^2+z^2}$) 
	of all cluster members for the same two clusters as in 
	Figure~\ref{2dclusters}. The most massive member is indicated by 
	the black circles. The boxes show the cluster center computed from 
	the MST. The line gives the trajectory for a particle moving with 
	an rms velocity of $\sqrt{3}c_{\rm s} = 0.35$ km s$^{-1}$. }
\label{rm6cluster} 
\end{figure}

Rather than relying solely on projected views, we can also consider
the full 3D picture at once. Figure \ref{rm6cluster} shows the root-mean-squared 
three-dimensional trajectory of the cluster members as a function 
of time for the same two clusters as shown in Figure~\ref{2dclusters}.
In this figure, the most massive cluster member's 3D position
is shown as a large circle at each time step, while other cluster
members are indicated by small coloured plusses.  At time steps when
the stars meet our criteria for a cluster, a proxy for the 3D cluster center 
is shown by the square.
In the upper panel, the stars only meet the MST cluster definition
for a short time, while in the bottom panel, the cluster
definition is satisfied for a longer period of time, although still
less than half of the time in which stars have formed.

Several important points are quickly apparent from these figures. First, the 
most massive member typically forms early;  
see Section~\ref{sec_star_times}.
Additional protostars which are eventually included in the MST-defined cluster form 
later, sometimes in close proximity to the most massive member and sometimes 
further away, migrating towards the position of the most massive 
member. Second, the motion of the most massive member is 
$\lesssim \sqrt{3}c_s$ and is relatively constant.
This is significant 
because dynamical interactions between cluster members may ``heat'' 
up the velocities as the cluster becomes virialized \citep{proszkow09}. 
Finally, although cluster members do appear to be dynamically interacting 
(e.g., top panel Figure \ref{rm6cluster}), the most massive member 
is already close to the MST-defined center often several $0.1$~Myr before 
a cluster is identified and before many close interactions occur. 

Some 2D-defined clusters contain members not associated in 3D space.
In Figure~\ref{rm6cluster}, this results in 
an offset of the three-dimensional center position relative to the most-massive
 member. Interestingly, these more distant members do not effect the 
conclusion that the protostars are primordially mass segregated. If anything, 
the random projected placement of these MST members would make the clusters 
appear less mass segregated than they actually are, by shifting
the apparent cluster centre away from its true location.  Observed 
mass-segregated clusters are thus likely to be 
more mass segregated than they appear 
due to the inclusion of unrelated stars due to chance alignments.

\subsection{Stellar Formation Times}
\label{sec_star_times}

\begin{figure*}
\begin{tabular}{cc}
\includegraphics[width=7.2cm]{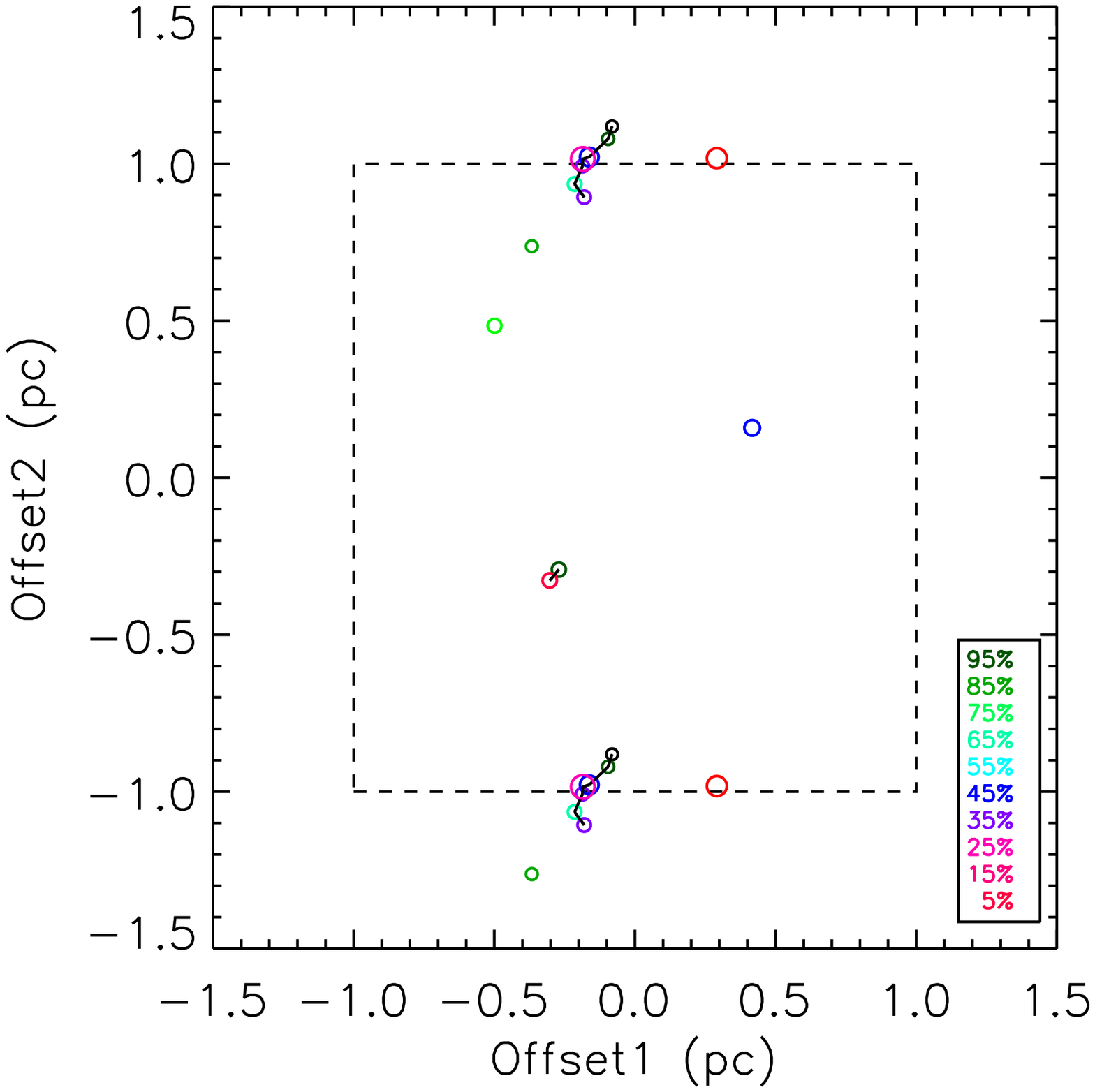} &
\includegraphics[width=7.2cm]{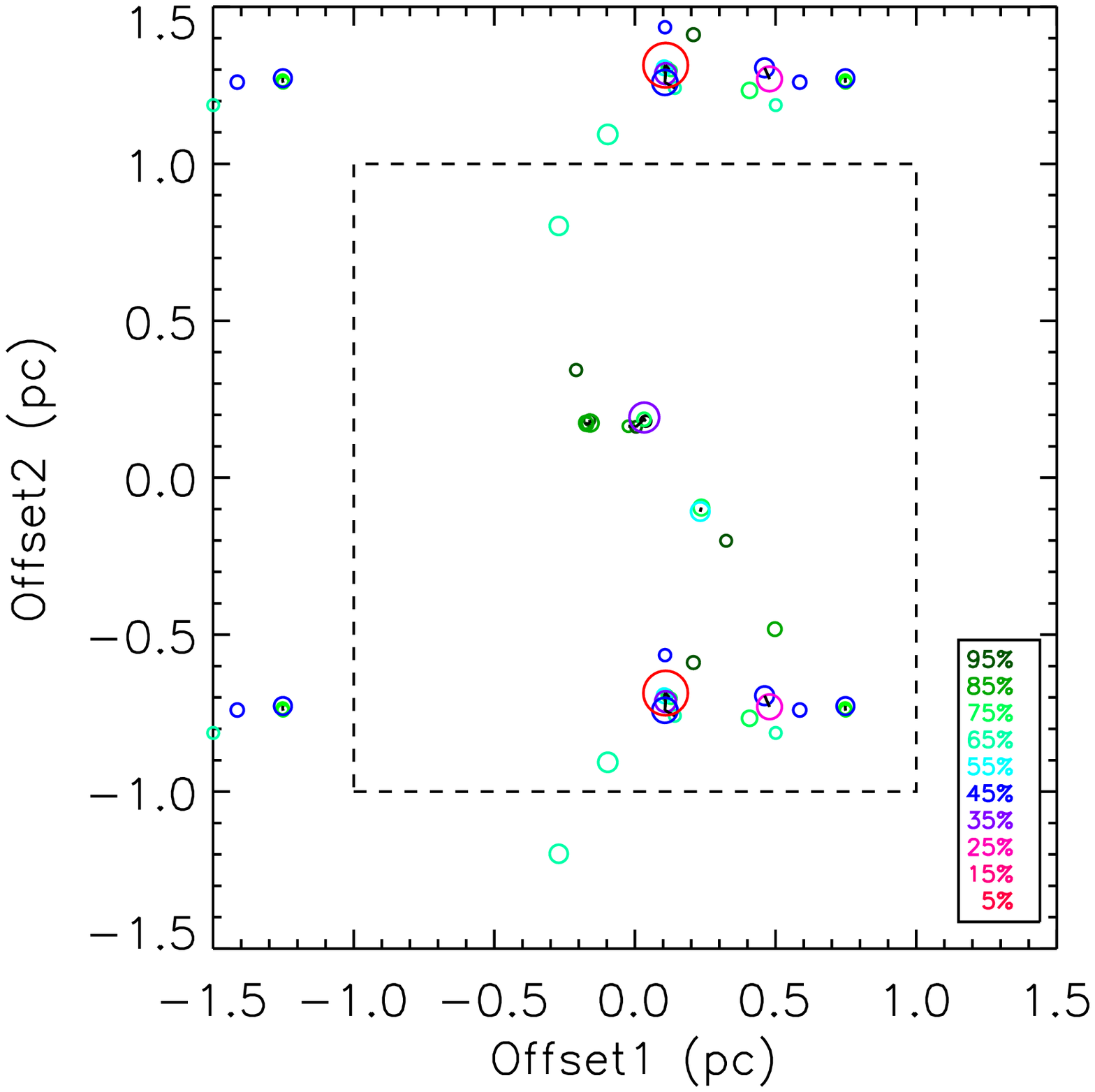} \\
\includegraphics[width=7.2cm]{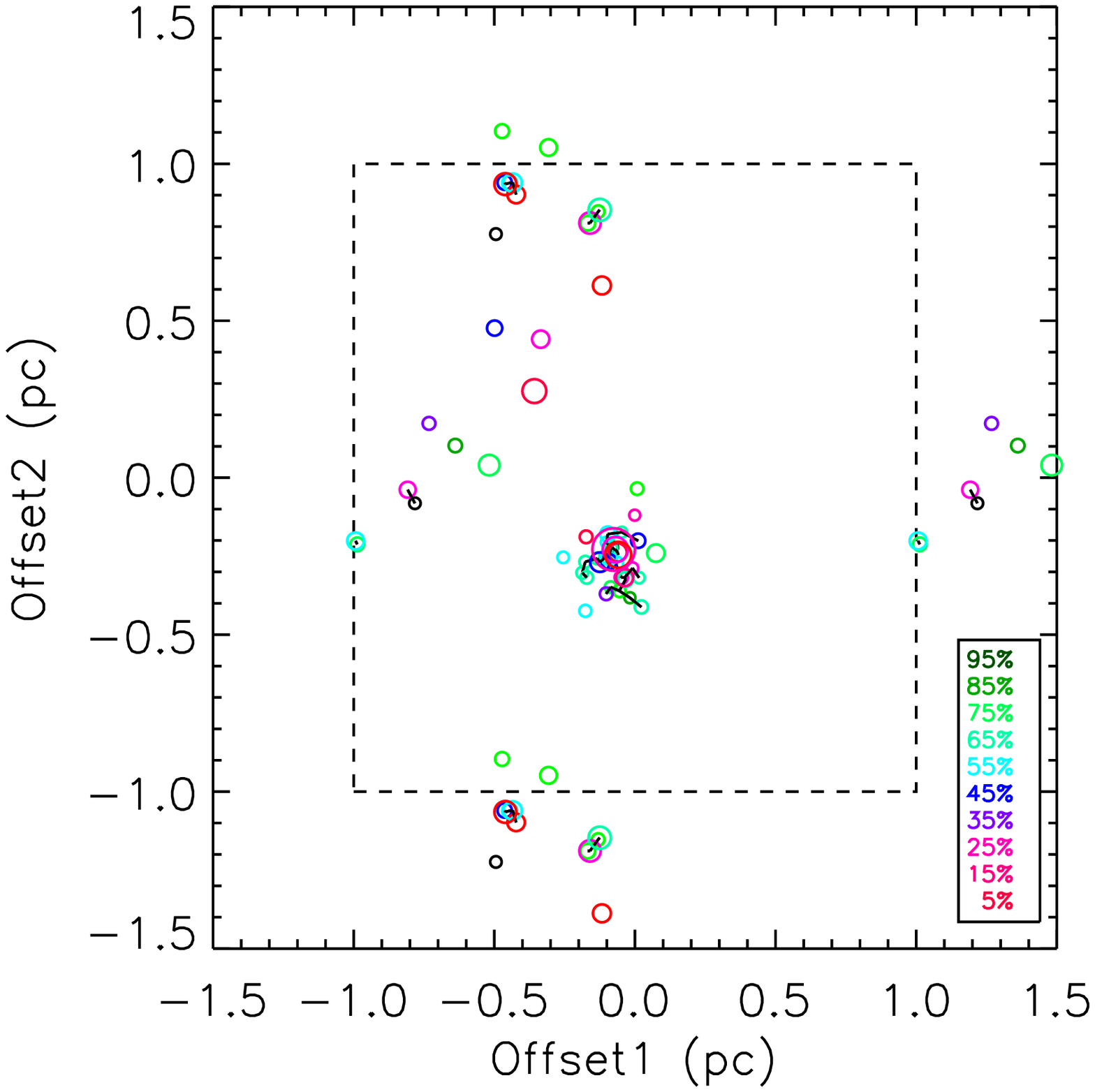} &
\includegraphics[width=7.2cm]{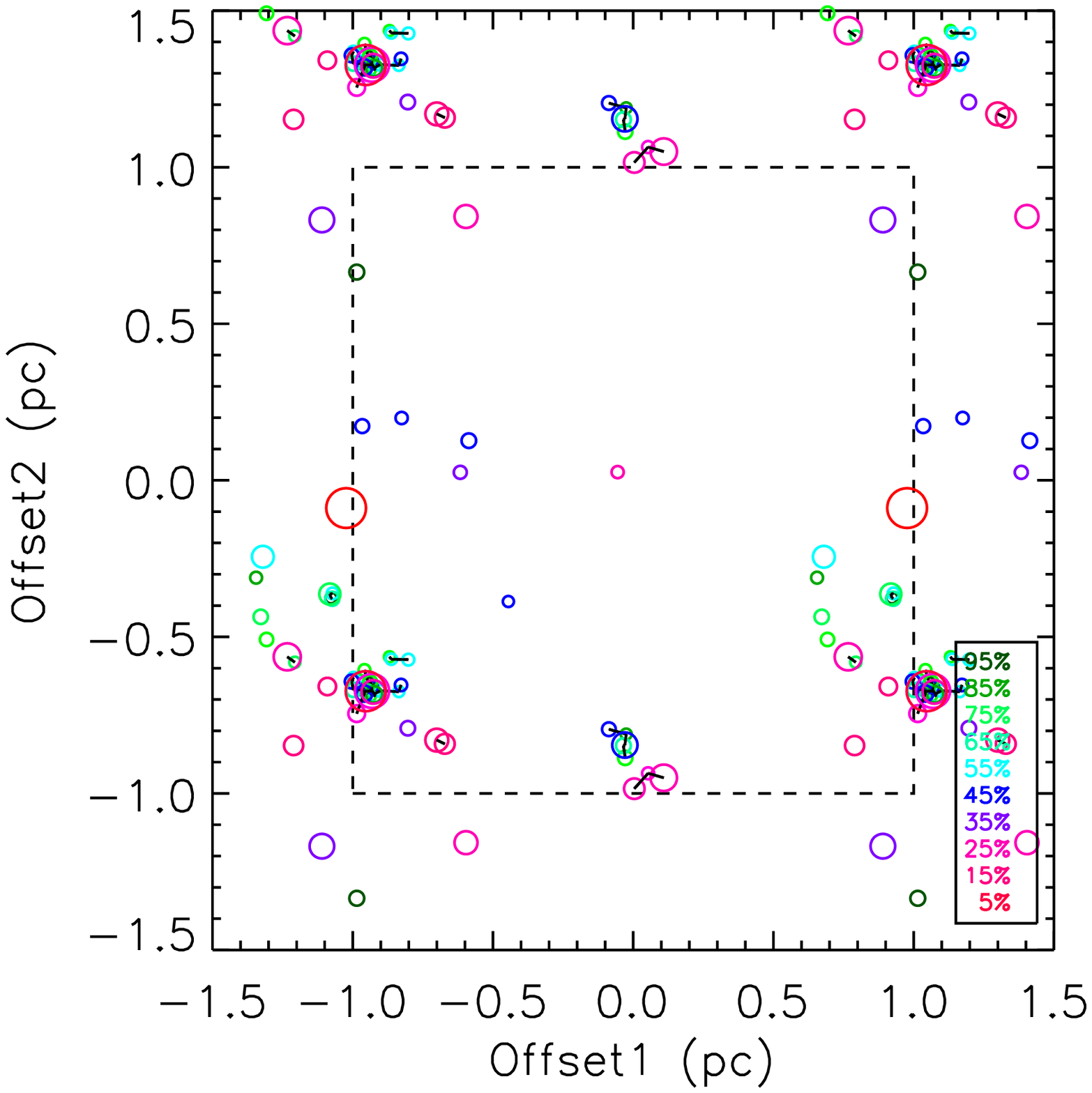} \\
\includegraphics[width=7.2cm]{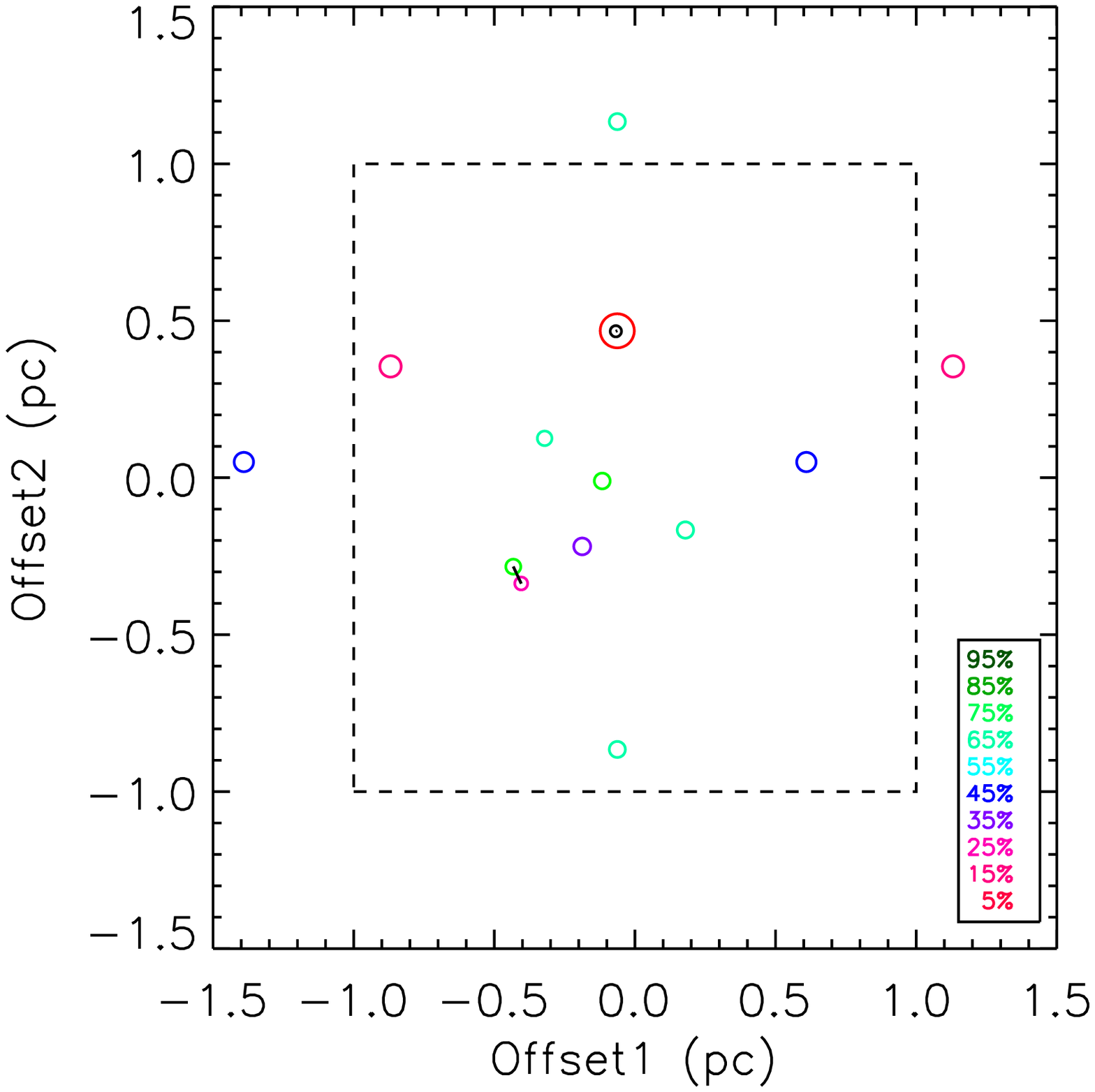} &
\includegraphics[width=7.2cm]{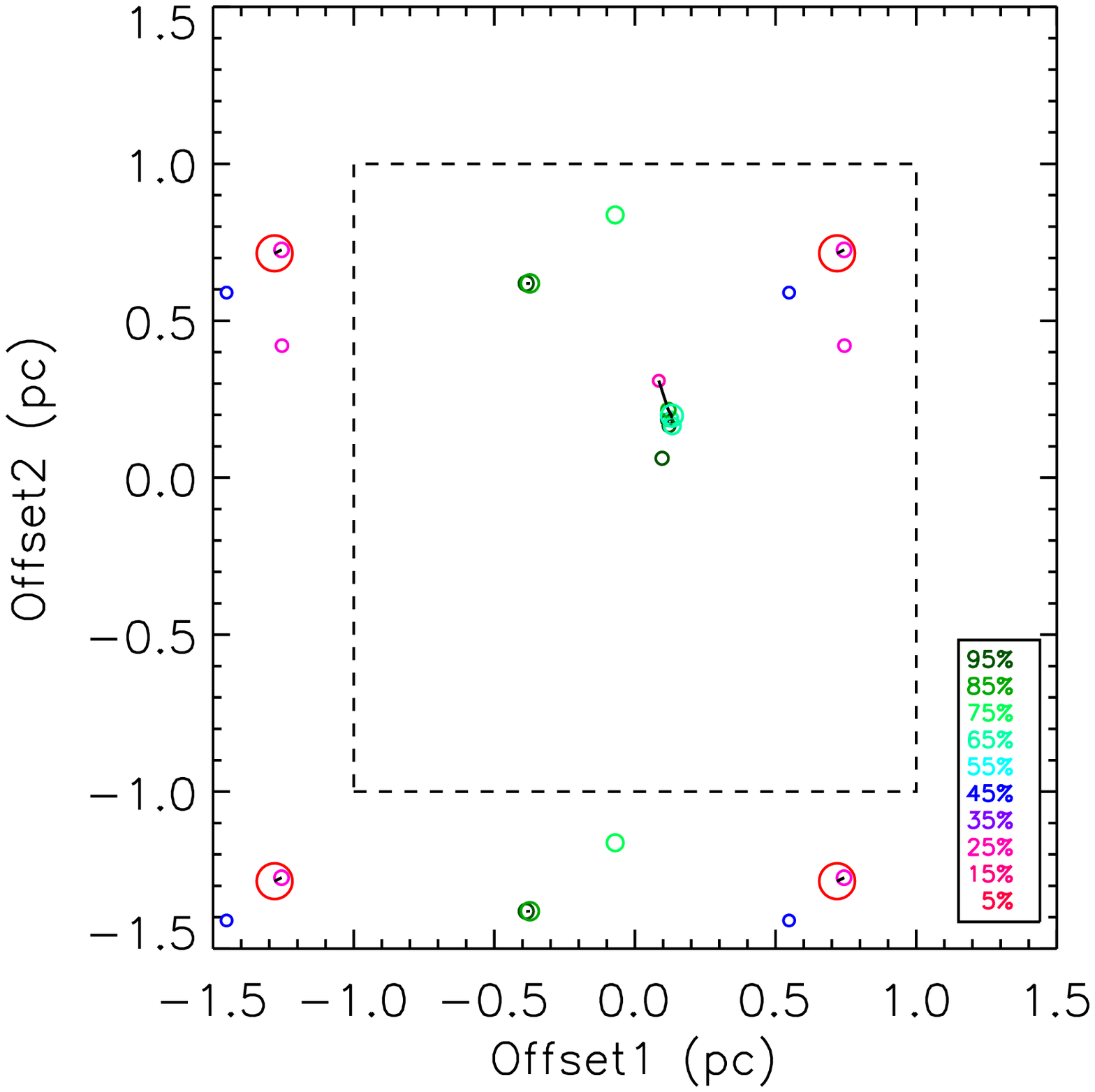}\\
\end{tabular}
\caption{The relative ages of stars formed in each of the simulations.
	For each simulation the $xy$ projection is shown at the final
	time step, with the same plotting convention for the stars and
	MST structure as previous figures, e.g., Figure~\ref{fig_mst_full}.
	The colour assigned to each star indicates when it formed in 
	the simulation as a fraction of the time elapsed since the
	first star in the simulation formed 
	(see legend on bottom right); larger percentages 
	thus indicate younger (more recently formed) stars.  
	The simulations shown are, from
	left to right, top row: Rk34 and Rm4, middle row: Rm6 and Rm6s, 
	bottom row: Rm9, Rt20.  Rrt formed the fewest stars and is
	omitted for brevity.
	}
\label{fig_all_mst_colour}
\end{figure*}

As noted in Section~\ref{sec_dyn}, the most  
massive member of each cluster tends to form early
in the simulation.  
Figure~\ref{fig_all_mst_colour} illustrates 
the relative formation times of the stars in the simulation, showing
the final time step of each simulation
except Rrt (excluded since it forms the fewest stars), viewed in
the $xy$ projection.  The stars are shown as the coloured circles,
with the circle size scaling with mass, and the MST structure 
is shown with the black lines.  The
{\it colours} assigned to each star represent how early in the 
simulation they first formed: red and purple indicate earlier 
times, while green and light blue indicate later times,
all calculated as the fraction of time elapsed since the
first star formed in the simulation.
What is immediately apparent from this figure is that the most
massive stars all {\it started} forming early in the simulations --
less than one third of the time since the first star formed.
Some low mass stars also formed early, but other low mass stars
continue to form throughout the simulation, as shown by the 
blue and green circles.
In Rm6 and Rm6s, where
MST clusters of more than ten members were identified using the best 
fit \Lcrns, the most massive star formed is a member of a cluster in
at least one projection.  This trend does not extend much further
in the mass-ranking, however, in several instances 
somewhat massive stars appear to be associated with smaller
(N$\le$10 member) groups for at least some time steps.  We emphasize
that although the massive stars may start forming early, this does
not imply that they also stop forming early.
Since there is no stellar feedback included in the simulation, which 
could reduce or halt accretion, it is not surprising that stars 
continue to accrete. 
A longer formation time for higher mass stars is consistent with certain 
types of analytic accretion models in which accretion rates depend weakly 
or are independent of stellar mass 
\citep{BateBonnell05,Myers09,McKee10}. 
A tendency for more massive stars to begin forming early has also been 
noted in some previous simulations 
(e.g., \citealt{klessen01,bate03,bonnell04,maschberger10}).
Furthermore, \citet{smith09a} find that the dense cores in which
massive stars form out of tend form first as well.

\section{Comparison with Observations}\label{sec_comp_obs}

\begin{figure}
\includegraphics[width=3.3in]{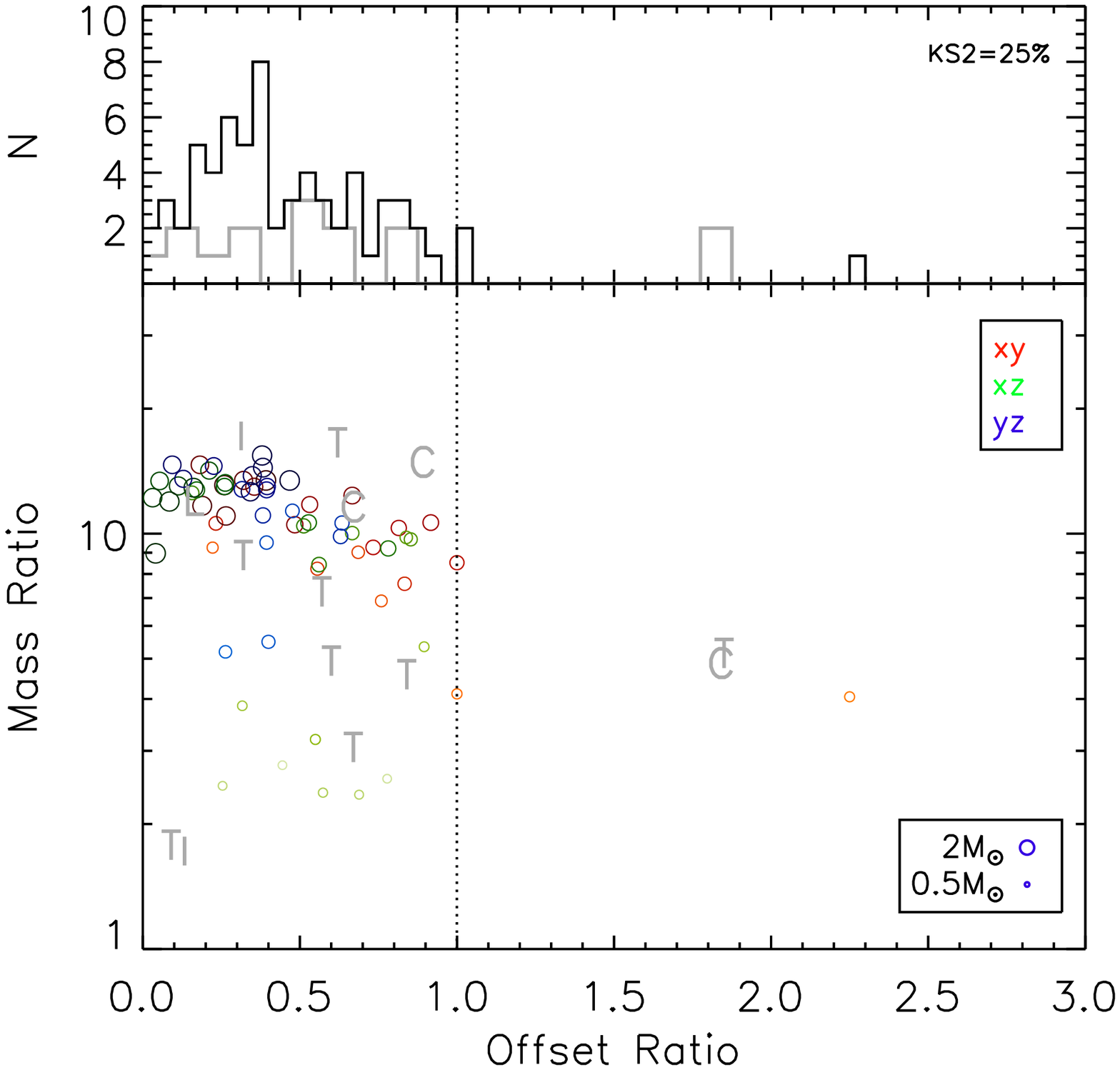}
\caption{The mass ratio and offset ratio for clusters identified in the
	Rm6 simulation, every 0.01~Myr, compared with observed cluster 
	properties.  Bottom panel: Red, green, and blue symbols show clusters
	identified for each projection,
	and lighter shades indicate earlier times in the simulation.
	The size of the circle indicates the mass of the most massive
	cluster member.  The grey symbols indicate the ratios measured
	in \citetalias{kirk11}.
	Top panel: The distribution of
	offset ratios in the simulation (black) and from 
	\citetalias{kirk11} (grey).
	The probability that the two distributions are drawn
	from the same parent sample is given in the upper right
	corner.  Randomly located most massive cluster members would
	lie around an offset ratio of 1 (dotted vertical line).
	}
\label{fig_Rm6_ratios}
\end{figure}

\citetalias{kirk11} found in their observational survey of young, small,
and nearby clusters, that the most massive cluster member tended
to be centrally located (offset ratios of $<$1), with no dependence
on other cluster properties, such as number of members, mass,
or mass ratio.  Figure~\ref{fig_Rm6_ratios} shows the mass and offset
ratios found in the Rm6 clusters compared with those in the
observational survey.  In the lower panel of Figure~\ref{fig_Rm6_ratios}, 
circles show the simulated
clusters, with the circle size varying with the mass of the most
massive cluster member, colour varying with projected view, and
shading varying with the time, sampled every 0.01~Myr.
The grey letters
indicate the observed cluster values in \citetalias{kirk11}
for {\bf T}aurus, {\bf L}upus3, {\bf C}haI, and {\bf I}C348.  
The simulated clusters
here clearly match the observations well -- nearly all have 
offset ratios less than one, and similar, though smaller range of
mass ratios.  There is no
trend in the offset ratio varying with the mass ratio or
most massive cluster member's mass.  The histogram in the
top panel of Figure~\ref{fig_Rm6_ratios} shows the distribution
of offset ratios for both the simulation (black line) and
observations (grey line) - both are similar.  A
KS test shows the probability that both observed and simulated
offset distributions are
drawn from the same parent sample is 25\% (top right corner); 
the two 
distributions are clearly consistent.

\begin{figure}
\includegraphics[width=3.3in]{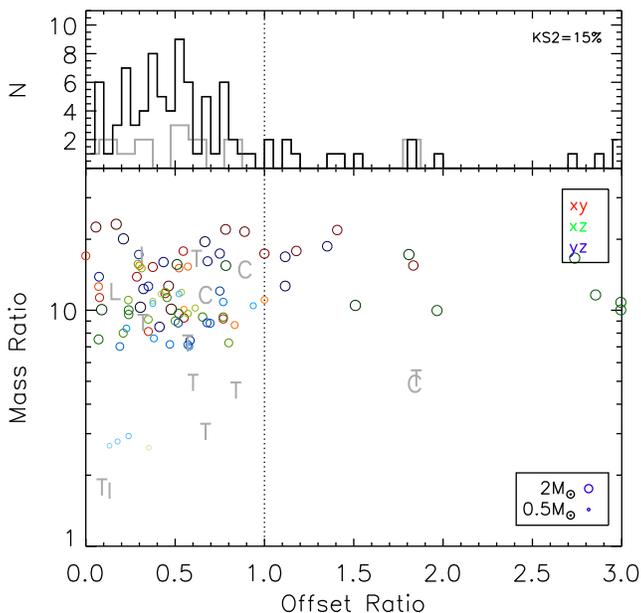}
\caption{The mass and offset ratios for clusters in the Rm6s simulation
	every 0.01~Myr.  See Figure~\ref{fig_Rm6_ratios} for the plotting
	conventions used.}
\label{fig_Rm6seed_ratios}
\end{figure}

Figure~\ref{fig_Rm6seed_ratios} shows the comparable results for
the Rm6s simulation, showing the same broad characteristics:
the majority of offset ratios are much less than 1, and mass
ratios around 10 to 20.  Rm6s appears to have generally slightly
higher mass ratios, and a broader distribution of offset ratios.
The small differences between the two thus reflect the difference 
between the two random turbulent seeds.

None of the other simulations had any clusters with more than 
10 members
using their best-fit \Lcrns.  Rm4 would have had 
a cluster identified at many time steps if a smaller minimum number
of members were adopted for the 
cluster definition (e.g., $> 5$ members).
All of the other simulations require both a smaller minimum 
number of members {\it and} a larger \Lcr in order for 
a cluster to be identified in more than one or two time steps
and a single projection.
We explore the effect of a more relaxed cluster definition in
Appendix~\ref{app_max_clust}, and find that 
the Rm6 and Rm6s clusters follow the same general
behaviour as seen in Figure~\ref{fig_Rm6_ratios} and \ref{fig_Rm6seed_ratios}, 
and `clusters' identified in the remaining simulations also
behave similarly (see Figure~\ref{fig_all_ratios_extra}).  
This implies that neither the specific  
cluster definition used nor the initial conditions has a 
strong influence on our results. 
At least in the regime of small-cluster
formation, our results suggest an interesting conclusion.  Changing the
initial conditions in the simulation from the fiducial
values which best match observations, to a larger or smaller Mach number,
higher temperature, smaller turbulent driving scale, or including radiative
effects has a significant impact on the number of stars formed, and their
general clustering properties (stellar density, visual appearance,
etc).  Despite this, the most massive stars tend to form in what will
become the centre of the cluster, suggesting that the mechanism 
for this is broadly applicable.

We note that the degree of mass segregation in local star-forming regions appears to be somewhat 
sensitive to how mass segregation is defined. For example, \citet{parker11} and \citet{parker12b} 
do not find mass segregation in Taurus and $\rho-$Ophiuchi, respectively. In fact, \citet{parker11} 
claim to find {\it inverse} mass segregation in Taurus. These results initially appear to be in conflict 
with our MST analysis and the results of \citetalias{kirk11}, who also analyze Taurus, but the 
difference can be understood by comparing the two samples. \citet{parker11} include all the stars in 
their analysis, essentially assuming that the entire region behaves as one cluster, while 
\citetalias{kirk11} make a point of identifying groupings within the sample, i.e., small-N analogs 
of higher mass clusters. We assert that subdivision is natural in regions such as Taurus, where star 
formation is distributed and stars in different areas appear to evolve with little knowledge of other subgroups.


\section{Conclusions}\label{conclusions}
Recent observations have shown that small stellar groupings
of $\sim 10 - 40$ members and only 1-2~Myr old, have a
centrally-located most massive member \citep{kirk11}, reminiscent of 
the mass segregation which is often claimed to be observed in
larger, denser stellar clusters such as the Orion Nebula Cluster
\citep{Hillenbrand98}.  In both cases, this leads to the larger
question of whether such mass segregation is primordial, or if 
it is a result of dynamical interactions between cluster members
early in the cluster's history.  
For large clusters, there
is also debate about whether observational biases (e.g., from stellar 
crowding in the cluster centre) causes some or all of the apparent 
mass segregation \citep{Ascenso09}.
Even if the mass segregation is real, some amount of dynamical
evolution is likely for larger clusters, since they tend to
be both older and have smaller stellar interaction timescales.
Observational bias is not present, however, for the small, sparse
clusters in \citet{kirk11}, and here we verify that dynamical evolution
is also unlikely to significantly impact or enhance mass segregation once a cluster has been identified.

Most previous simulations have focussed on higher density 
systems, where dynamical interactions are expected to play an important role in producing mass segregation. 
We analyze a suite of numerical simulations with varying
initial conditions which form small
clustered systems, in order to investigate the early dynamics
of such systems.  
We examine the effect of variations in the Mach number,
temperature, and driving scale on clustering, including one case 
which includes radiative transfer.
We analyze the small stellar clusters formed in these simulations
in the same manner as applied to the observations.  At each time step,
we ``observe'' the simulation from three viewing angles, removing
the lower-mass companions in tight pairs which would be unresolved
and any stars with masses below the completeness limit in
\citet{kirk11}.  We identify clusters using the minimal spanning
tree (MST) formalism \citep[e.g.,][]{gutermuth09}, and then apply the
same analyses as in \citet{kirk11}.  Ours is the first study
which applies the MST analysis to simulations in an identical manner 
to observational
work, although \citet{maschberger10} and \citet{girichidis12b} do
apply an MST without observational cutoffs for their analysis.

Only the fiducial simulation and a second random realization of the
same initial conditions result in clusters satisfying the
observational requirements (more than 10 members connected in an
MST by branch lengths of less than the locally-measured \Lcrns).
Relaxing the observational requirements allows clusters in the other
simulations to also be identified.  In all cases, we find the simulated
clusters mimic the observations: the most massive cluster member
tends to be centrally located.  We furthermore find that the 
most massive cluster member's position within the cluster shows
little to no evolution with time.
`Mass segregation', such as it can be observed
in such small clusters, is primordial in the simulations.  This agrees well
with various other lines of evidence suggesting that dynamics do
not play a significant role in small, young, stellar clusters. 
For example, observations of low-mass cores show that core 
relative motions are small compared to the local turbulent velocity
\citep{kirk07, andre07, roso08}.  Furthermore, dense core gas and envelope gas
(as traced by, e.g., N$_2$H$^+$ and C$^{18}$O, respectively) 
also have similar line-of-sight
velocities \citep{kirk10}. Both findings suggest that
stars are not formed with initially ballistic motions relative to the
surrounding gas. 
Some turbulent simulations are able to reproduce the observed 
low core-to-core velocity dispersions \citep{Offner08b}. Such simulations also
demonstrate that protostellar velocities are initially subvirial relative 
to the gas \citep{Offner09b}. If stars inherit their motions from
the dense gas and, consequently, are slow-moving, then mass segregation
observed after 1-2~Myr in low-mass star forming regions is likely 
predominantly primordial.  
Observations of protostellar luminosities in local star forming 
regions find that brighter sources are preferentially located in higher 
protostellar density regions, which may indicate early mass segregation 
\citep{kryukova12}.
This conclusion could also be confirmed by examining the distribution of
core masses; assuming a one-to-one mapping between dense cores and protostars,
mass segregation should then also be detectable at the dense core stage.
Note, however, that in more clustered environments, individual dense cores become
difficult to separate, and objects identified are likely to
be multiple dense cores blended together, despite the fact that these
composite objects may have a similar mass function to the IMF \citep{reid10,michel11}.

Finally, we note that there is a tendency for the most massive
cluster member to form relatively early in the simulation. 
\citet{maschberger10} found a similar trend in their analysis
of a numerical simulation of large cluster formation; there, they
note that the most massive cluster member forms early, within
subclusters that merge to form larger clustered systems. 
Having the most massive member starting to form early is perhaps
not unexpected: it has the most mass to accrete, and therefore
may need more time overall to do so \citep[e.g.,][]{Myers09}.

Unlike the simulations we analyze, stars forming in high-stellar
density regions likely undergo many interactions on a short time
scale. Thus, our conclusions do not apply to massive star forming
regions, which may evolve quite differently.

\appendix
\section{Critical Length Determination}
\label{app_critlen}

The analysis and results presented throughout this paper rely on
the definition of a cluster adopted.  Using the MST formalism,
there are two parameters which control the identification of
clusters: the critical length scale beyond which stars are
not connected to a cluster, \Lcrns, and the minimum
number of members to be classified as a cluster, $N_{min}$.
For the simulations we analyze, \Lcr has a greater influence
on the clusters identified, because all of the clusters
identified are small.  $N_{min}$ therefore cannot be raised much above
11 for clusters to be identified in any simulation, 
and $N_{min}$ also cannot
be lowered much below 11 before properties such as the cluster centre
become difficult to measure.  The effect of a lower $N_{min}$ is
examined in more detail in Appendix~\ref{app_max_clust}.
In this section, we focus on the uncertainties associated with
determining \Lcrns, and the impact of these on our results.

\begin{figure}
\includegraphics[width=8cm]{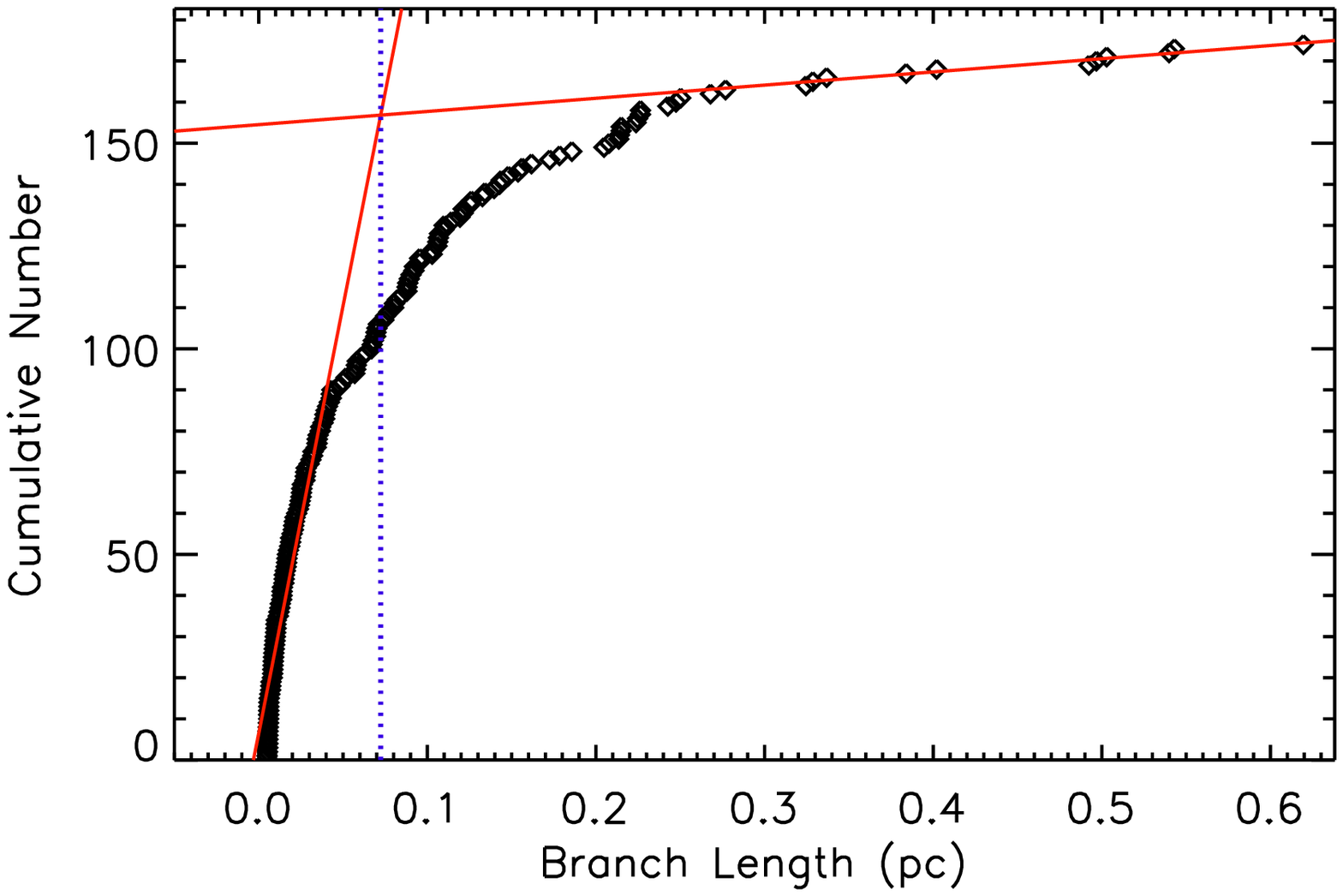}
\caption{The distribution of MST branch lengths for Rm6s.
	The red sold lines show the linear fits to the two
	edges of the distribution, and the blue dotted line
	shows the critical length measured, $\sim0.07$~pc.}
\label{fig_brlens_Rm6seed} 
\end{figure}

\begin{figure*}
\begin{tabular}{cc}
\includegraphics[width=8cm]{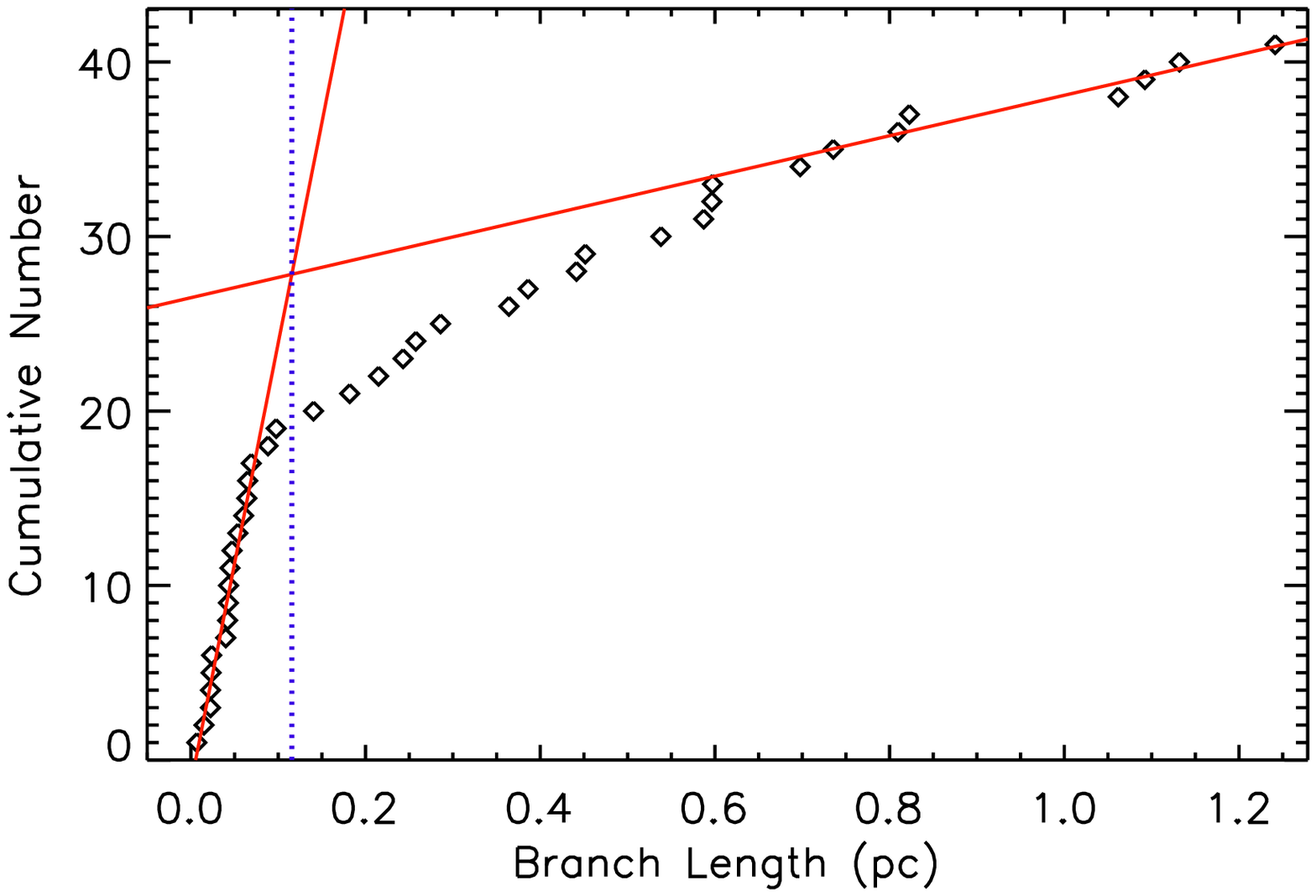} &
\includegraphics[width=8cm]{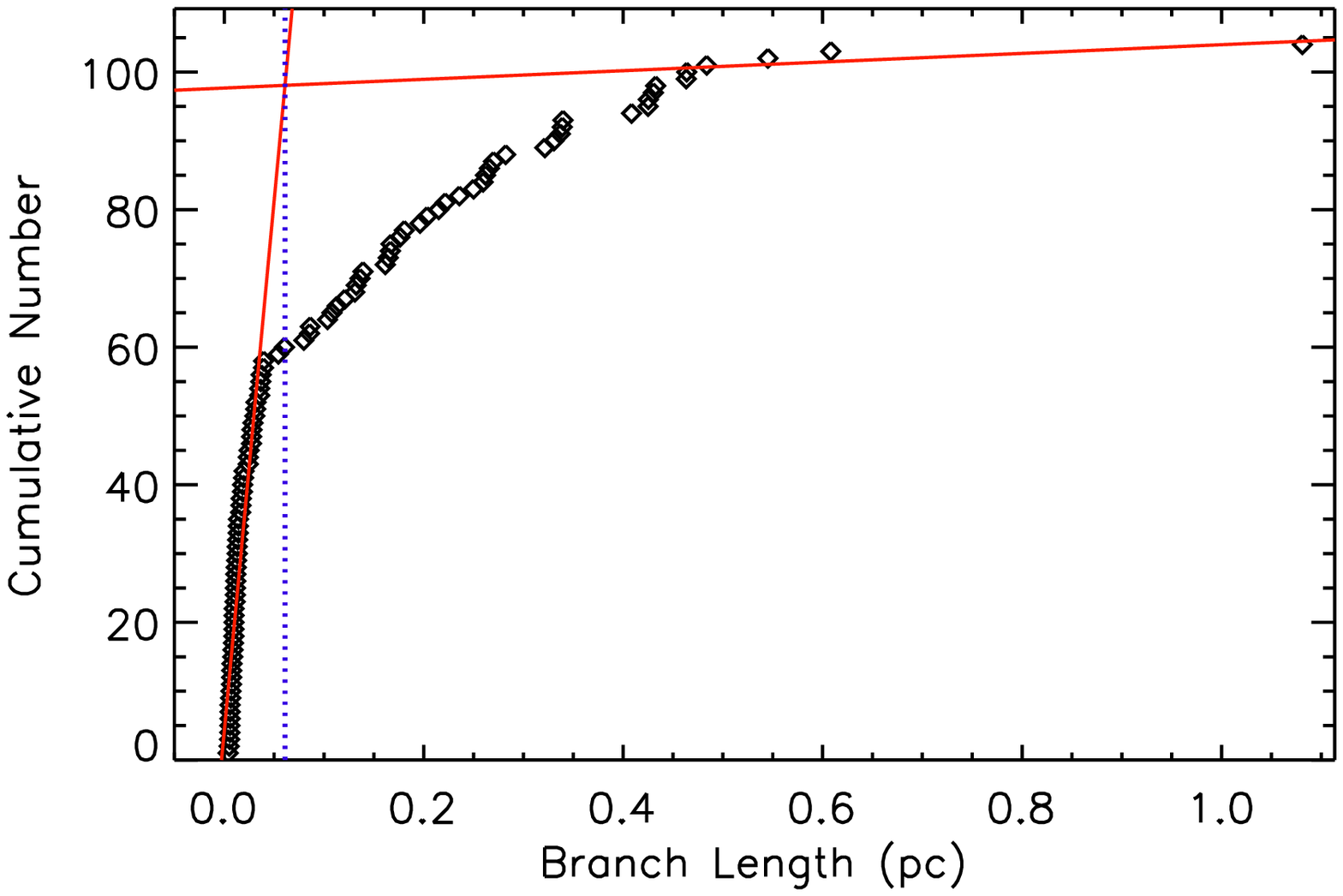}\\
\includegraphics[width=8cm]{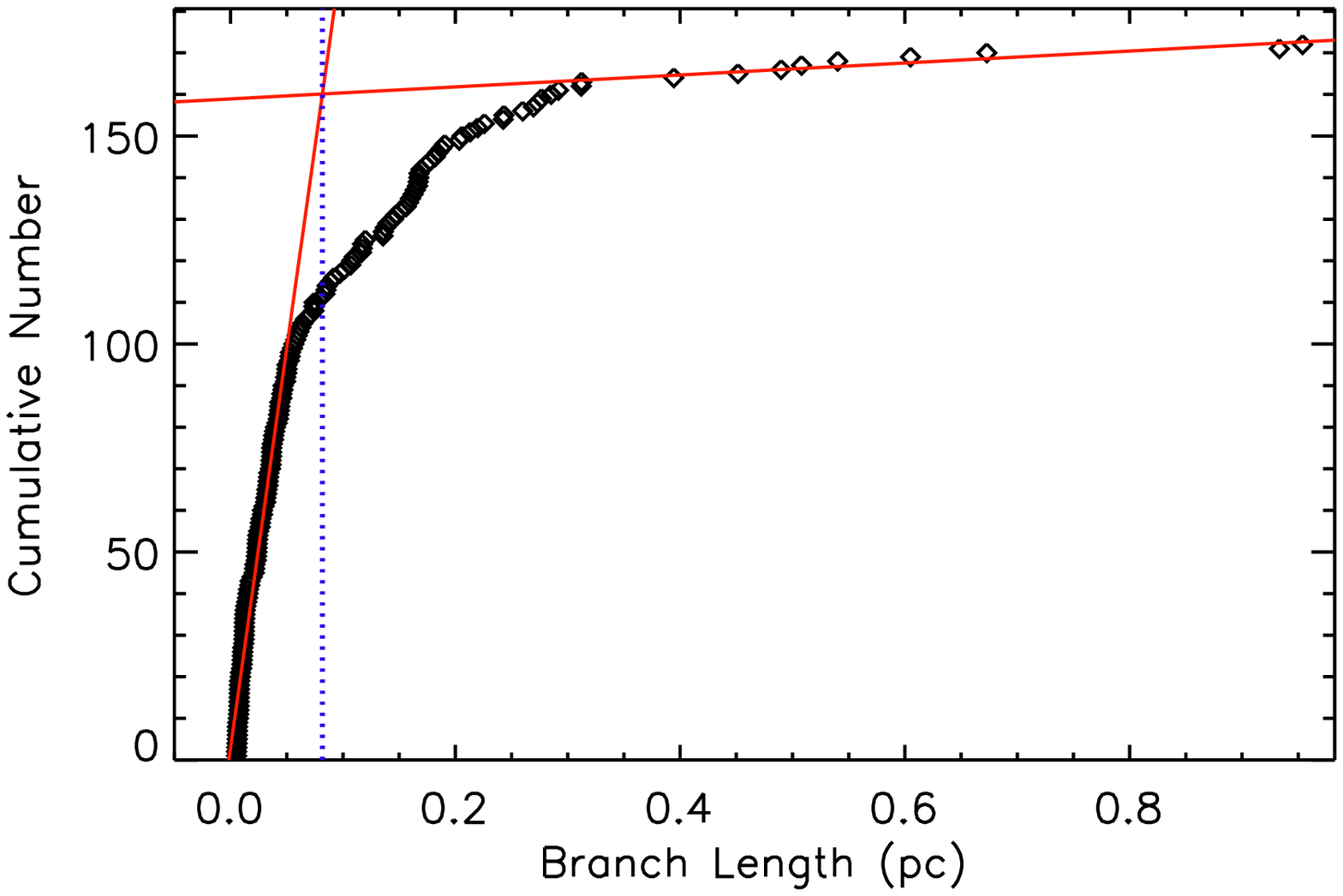} &
\includegraphics[width=8cm]{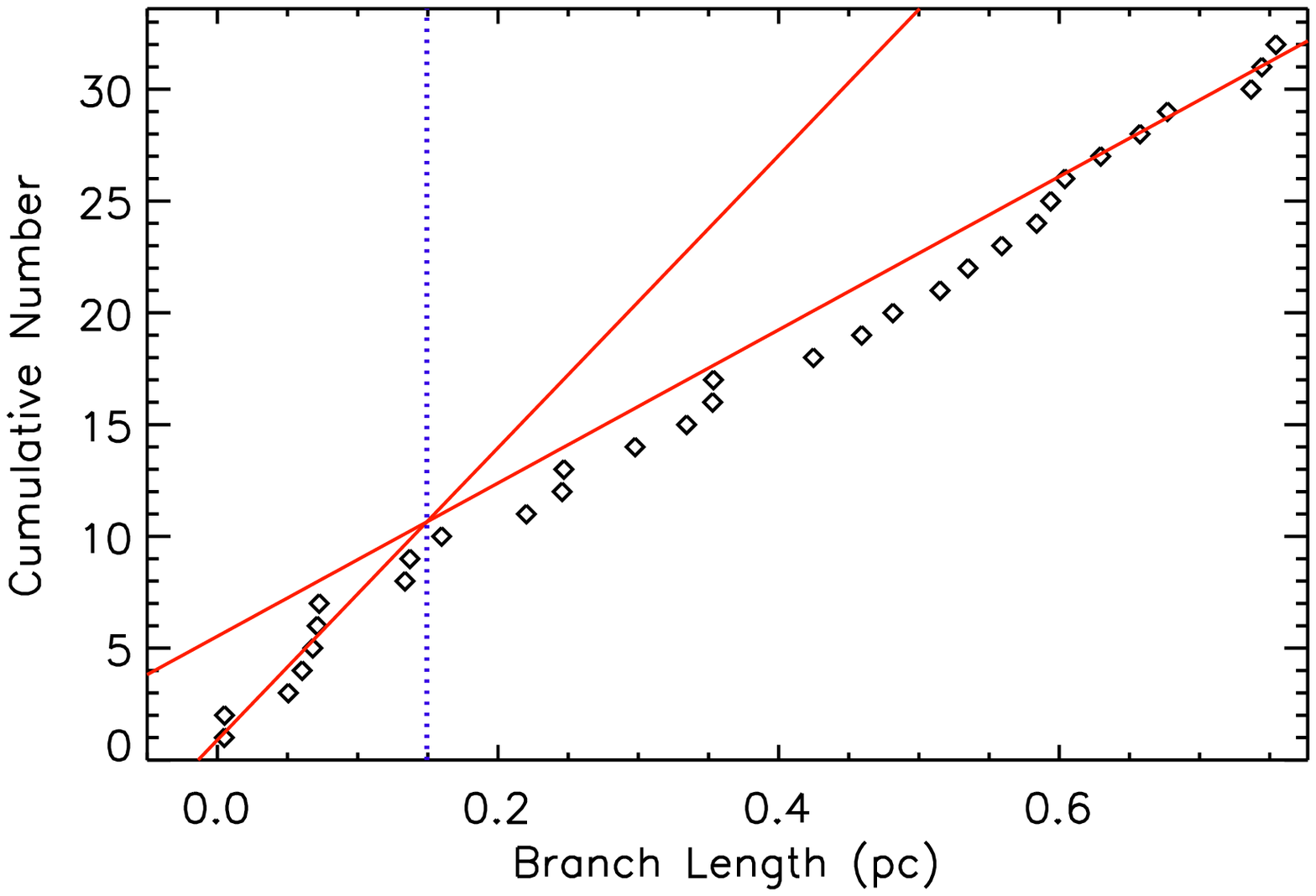}\\
\includegraphics[width=8cm]{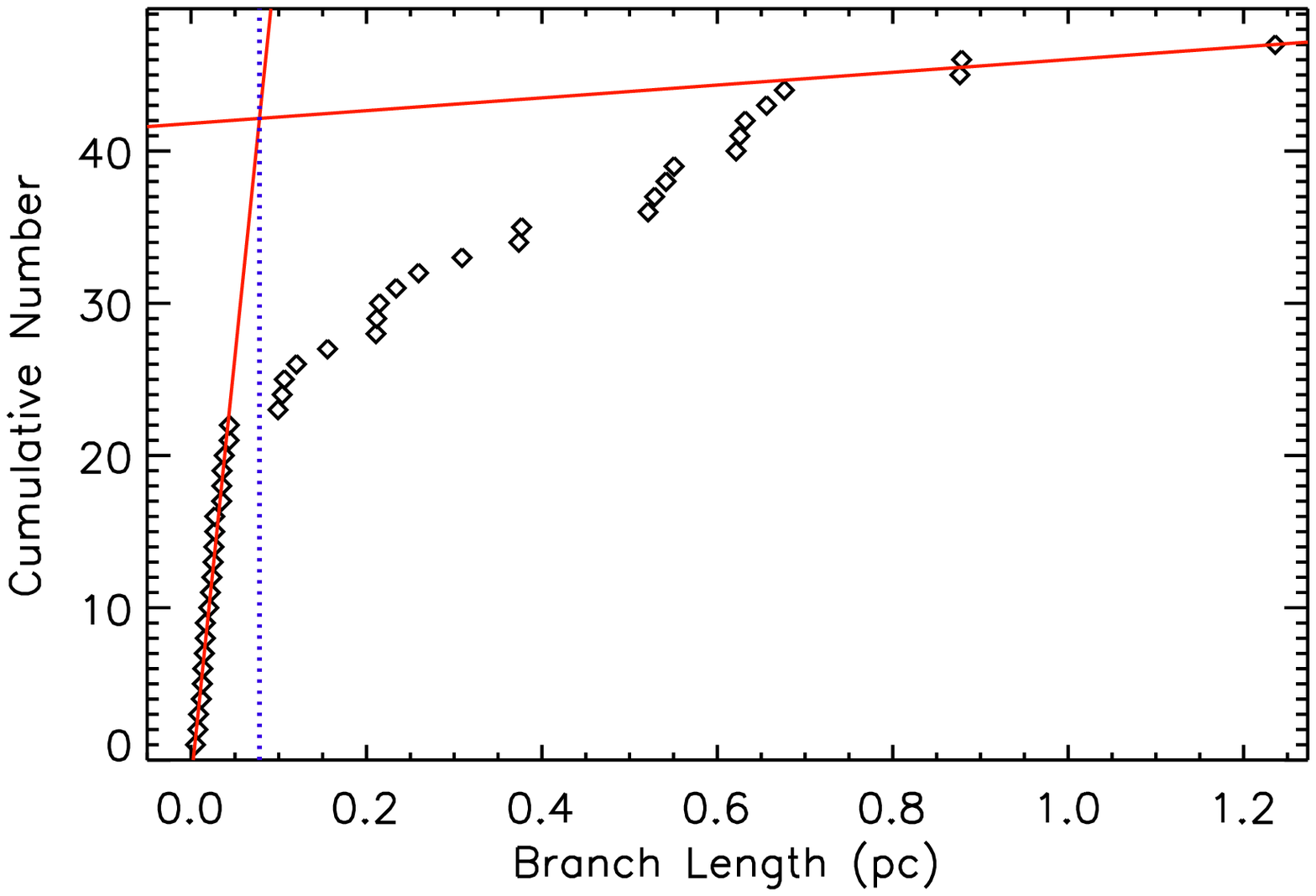}&
\includegraphics[width=8cm]{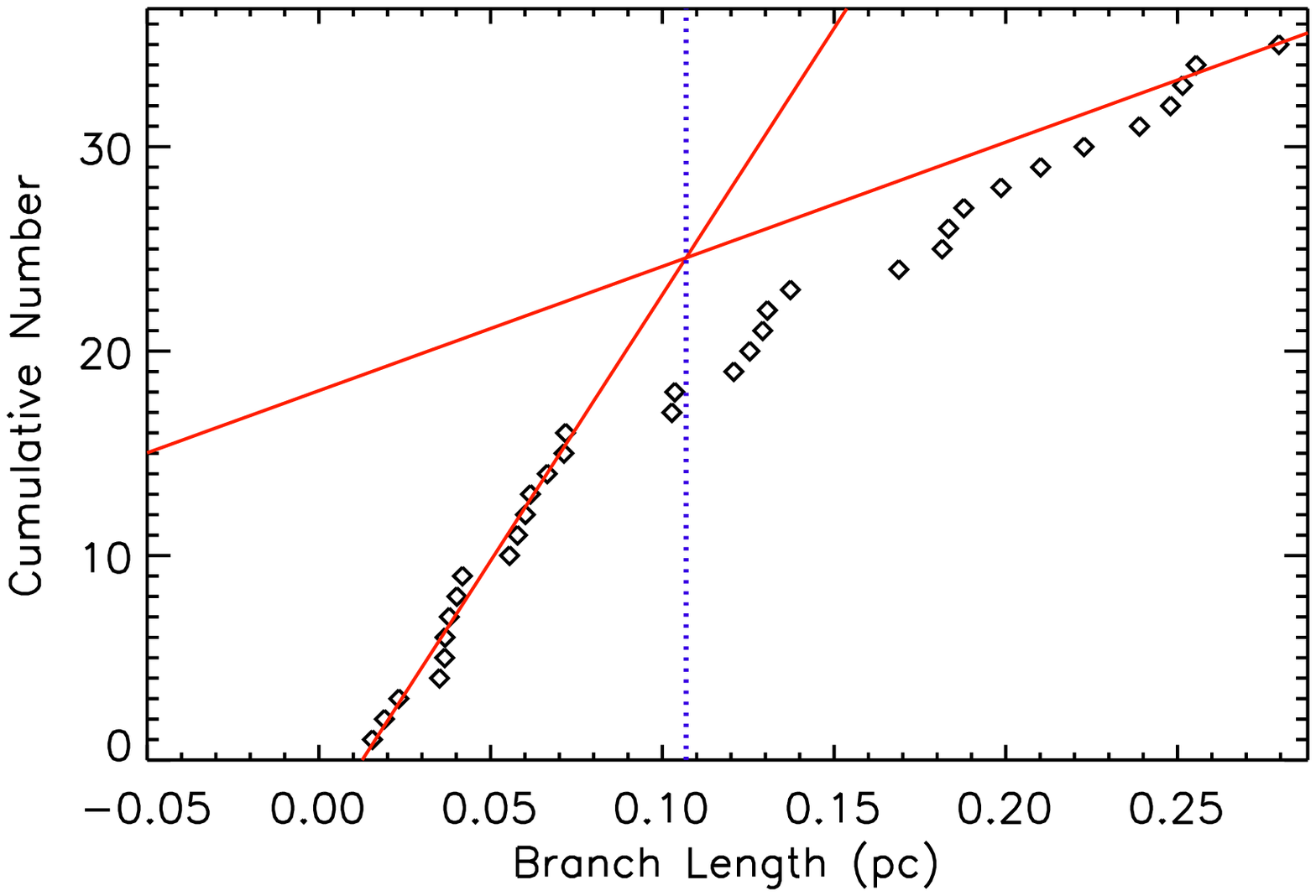}\\
\end{tabular}
\caption{The distribution of MST branch lengths for all of the remaining
	simulations.  Top row: Rk34, Rm4, middle row: Rm6, Rm9, and
	bottom row: Rt20, Rrt.  The red solid lines show the linear
	fits to the two edges of the branch length distribution,
	and the blue dotted line shows the critical length measured.
	The critical lengths are given in Table~\ref{tab_critlens}.}
\label{fig_brlens_all}
\end{figure*}

As discussed in Section~\ref{sec_mst}, we determined \Lcr following
the procedure used in \citetalias{gutermuth09} and \citetalias{kirk11}.  
\citetalias{gutermuth09} found that MSTs in clustered star-forming
regions tend to have a characteristic cumulative branch length
distribution: a sharp, roughly linear rise at small
lengths, followed by a turn-over, and a roughly linear
shallow rise at large branch lengths.  This is 
illustrated in Figure~\ref{fig_brlens_Rm6seed}, showing the 
cumulative branch length distribution for the Rm6s simulation at the 
final timestep.
\citetalias{gutermuth09} performed 
a linear fit to the steep and shallow slopes of the distribution, 
defining \Lcr as the 
intersection between these two best-fit lines.  This point gives an
approximation of the turn-over location in the cumulative branch
length distribution.
We followed the same procedure in our analysis;
the linear fits for Rm6s are shown in Figure~\ref{fig_brlens_Rm6seed}.
Note that while our full MSTs from the simulations contain the
nine replications of the initial simulated box (to account
for edge effects within the periodic boundaries), we took
care to remove any duplicate branches prior to fitting.
More accurate fits are obtained with a larger number
of branches, and so we included all three projections
at once for our critical length measurement.  Examination
of the branch length distribution for each projection
separately revealed similar critical lengths when there 
were sufficient points for a good fit.  Figure~\ref{fig_brlens_all}
shows the cumulative branch length distributions and \Lcr
determinations for the remaining six simulations.  
Qualitatively, many of the simulations have a 
markedly different {\it shape} to the cumulative branch
length distribution than is observed -- Rm6s appears the most
similar to observed distributions, while simulations such as Rm9
are notably different, with too-steep slopes at both small and
large branch lengths.  This difference may be partially attributable
to the small number of branches in several of the simulations, or it 
may be indicative of differences in the clustering properties.

\subsection{Uncertainties}
For cumulative branch length distributions where the slope of the steeper 
slope is not extreme, such as in Rm6s or Rm6, \citetalias{kirk11}
found that the uncertainty in \Lcr is roughly 10\%.  In the
case of the steepest branch length distributions (e.g., Rk34),
the uncertainty in \Lcr is likely larger.  Examining the
cumulative branch length distributions, however, also suggests
that a change of that magnitude in \Lcr will have a relatively
minor impact on the clusters identified.  Moving the dashed
blue \Lcr line in Figures~\ref{fig_brlens_Rm6seed} and 
\ref{fig_brlens_all} by 10\% or even 20\% does not result in
a large change in the cumulative number (vertical axis)
of MST branches which could be connecting stars in a 
cluster.  This is an upper limit to the number of stars whose
cluster membership would start (end) with the increase (decrease)
in \Lcrns; some of the stars indicated by these points 
would be associated with groupings of stars which
are too small to meet the minimum cluster size.  Only a few
stars at a cluster's periphery are therefore likely to change
their cluster membership status within reasonable variations
of \Lcrns.  \citetalias{kirk11} found similar results in their
observational survey.  

\subsection{Time Evolution}
The cumulative branch length distributions shown in Figures
\ref{fig_brlens_Rm6seed} and \ref{fig_brlens_all} are
all from the final time step that each simulation was run.
In principle, a similar analysis should be made at each time
step for each simulation.
Where the number of branches
was sufficiently large to perform accurate fits, we searched for evidence of 
time evolution of \Lcr or the shape of the cumulative branch 
length distribution.
For Rm9 and Rrt, too few sources were ever
present to allow for a good determination of \Lcrns, as noted
in Table~\ref{tab_critlens}.  For Rk34 and Rt20,
the paucity of sources, particularly at earlier time steps,
was sufficient to lead to a $>$10\% scatter in \Lcrns, although
the general shape of the cumulative branch length distribution does not
change significantly after the first few time steps examined.
For Rm4, Rm6, and Rm6s, which had the greatest
number of sources, the overall shape of the cumulative branch
length distribution appears to change little over time.  There
is some variation in the values of the \Lcr fit, but no 
definitive evidence for systematic increase or decrease in \Lcr 
with time.  At the earliest times when the number of sources
is smallest, both the shape of the cumulative branch length
distribution and the fit \Lcr tends to be more variable,
often appearing more similar to, e.g., the Rm9 distribution
in Figure~\ref{fig_brlens_all}.  At the time steps examined,
\Lcr for Rm6 tended to lie between 0.06~pc and 0.08~pc,
although in a few cases, the best fit values had more extreme
values (between 0.05~pc and 0.09~pc to 0.1~pc).
Rm6s showed a slightly larger scatter in \Lcr values, but similarly 
most values were between 0.06~pc and 0.08~pc.

Given these findings, we argue it is reasonable to adopt a single,
constant value of \Lcr for each simulation; there is no strong
evidence for evolution of \Lcr with time which might 
potentially bias our results.  In Appendix~\ref{app_max_clust}
we further demonstrate that our results are robust to the
precise \Lcr adopted; we find similar results even when \Lcr
is increased beyond the uncertainty discussed here.

\subsection{Maximal Cluster Membership}
\label{app_max_clust}
To further test the effects of our cluster definition
on our results, we re-run the analysis shown in
Section~\ref{sec_comp_obs} with a more inclusive cluster definition.
We change our cluster definition to a minimum of 6 stars (from 11).
When a cluster has few stars, the centre position is poorly defined,
generating
a larger scatter in the offset ratios measured.
We also increase \Lcr to the maximum allowable before all
stars in the simulation are connected in a single cluster,
0.15~pc for Rm4, Rm6, Rm6s, and Rrt, 0.3~pc for Rt20, and 0.5~pc for
Rk34 and Rm9; these are at least 50\% larger than the best fit values for 
each.
[Note that
since the simulations have periodic boundary conditions, defining
the cluster centre would become problematic if all stars belong to
the same cluster.] 

\begin{figure}
\includegraphics[width=8cm]{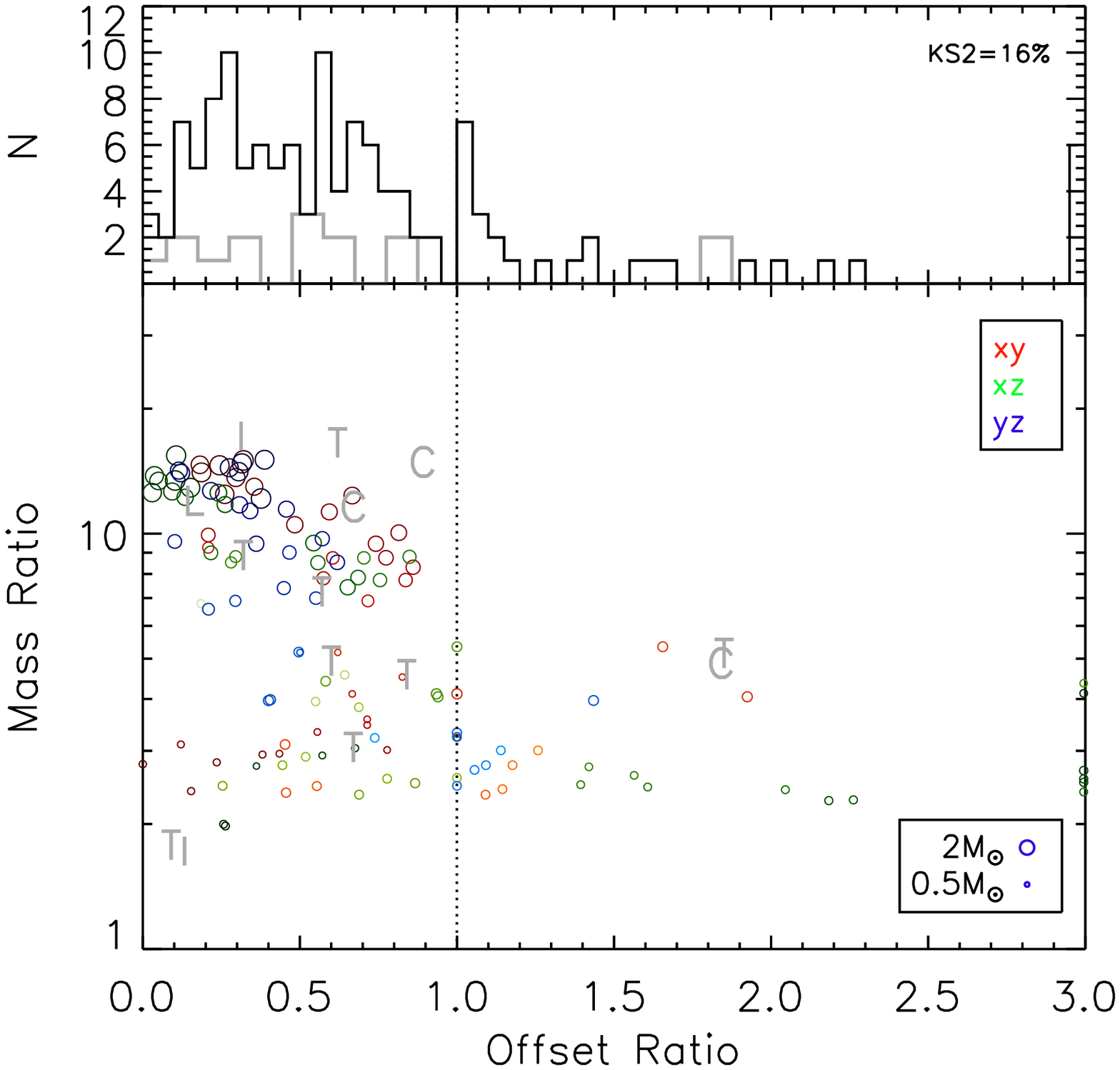}
\caption{The mass and offset ratio for clusters identified in the
	Rm6 simulation, with the value of $L_{crit}$ increased to
	0.15~pc from 0.07~pc, and the minimum number of members
	required to be classified as a cluster reduced to 6 (from 11).
	Clusters with offset ratios of more than 3 are all plotted
	as having a value of precisely 3.
	The same trends are followed
	as in Figure~\ref{fig_Rm6_ratios} are found for this relaxed
	cluster definition.  See Figure~\ref{fig_Rm6_ratios} for
	the plotting conventions used.}
\label{fig_Rm6_ratios_extra}
\end{figure}

\begin{figure*}
\begin{tabular}{cc}
\includegraphics[width=7.0cm]{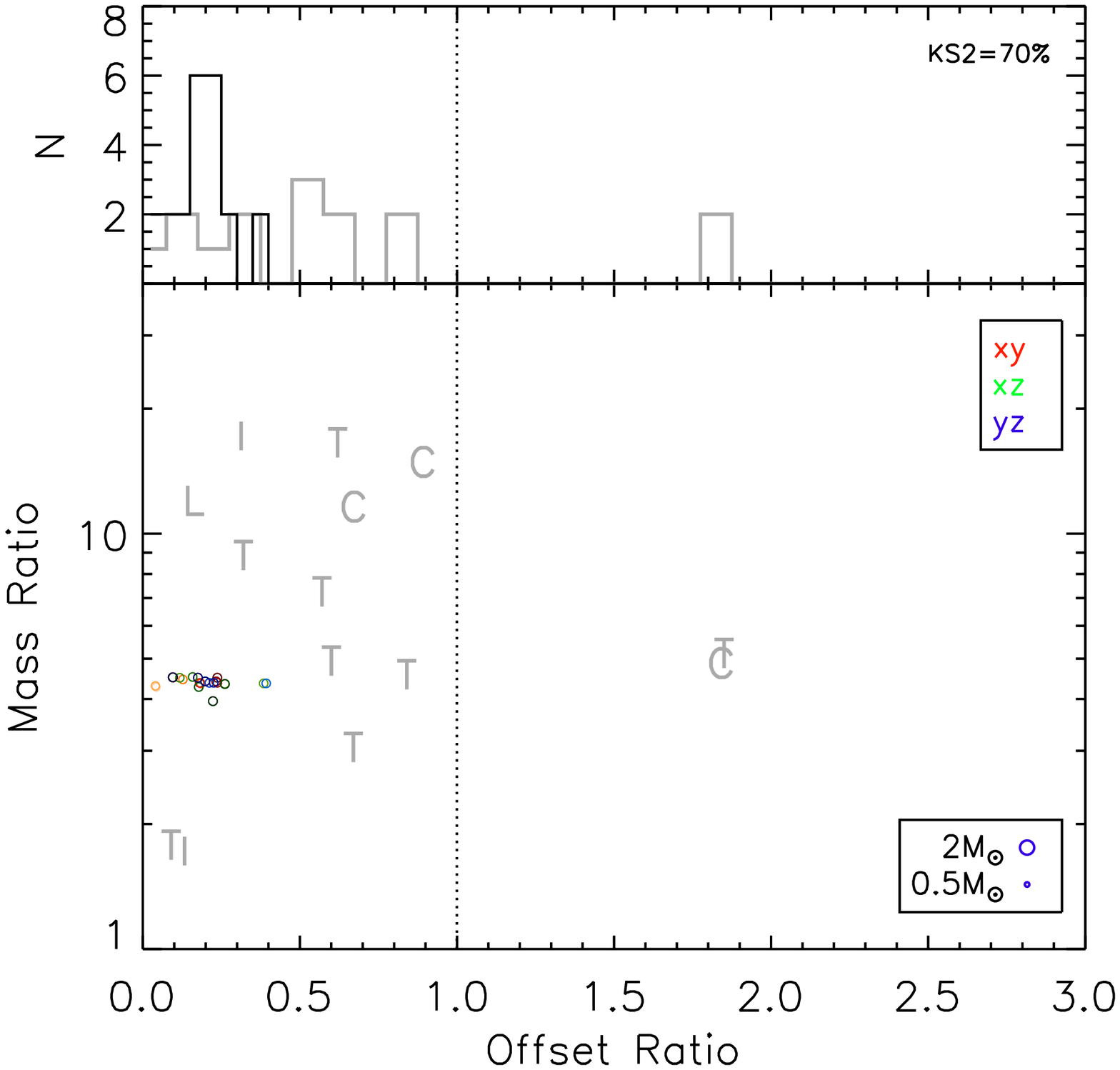} &
\includegraphics[width=7.0cm]{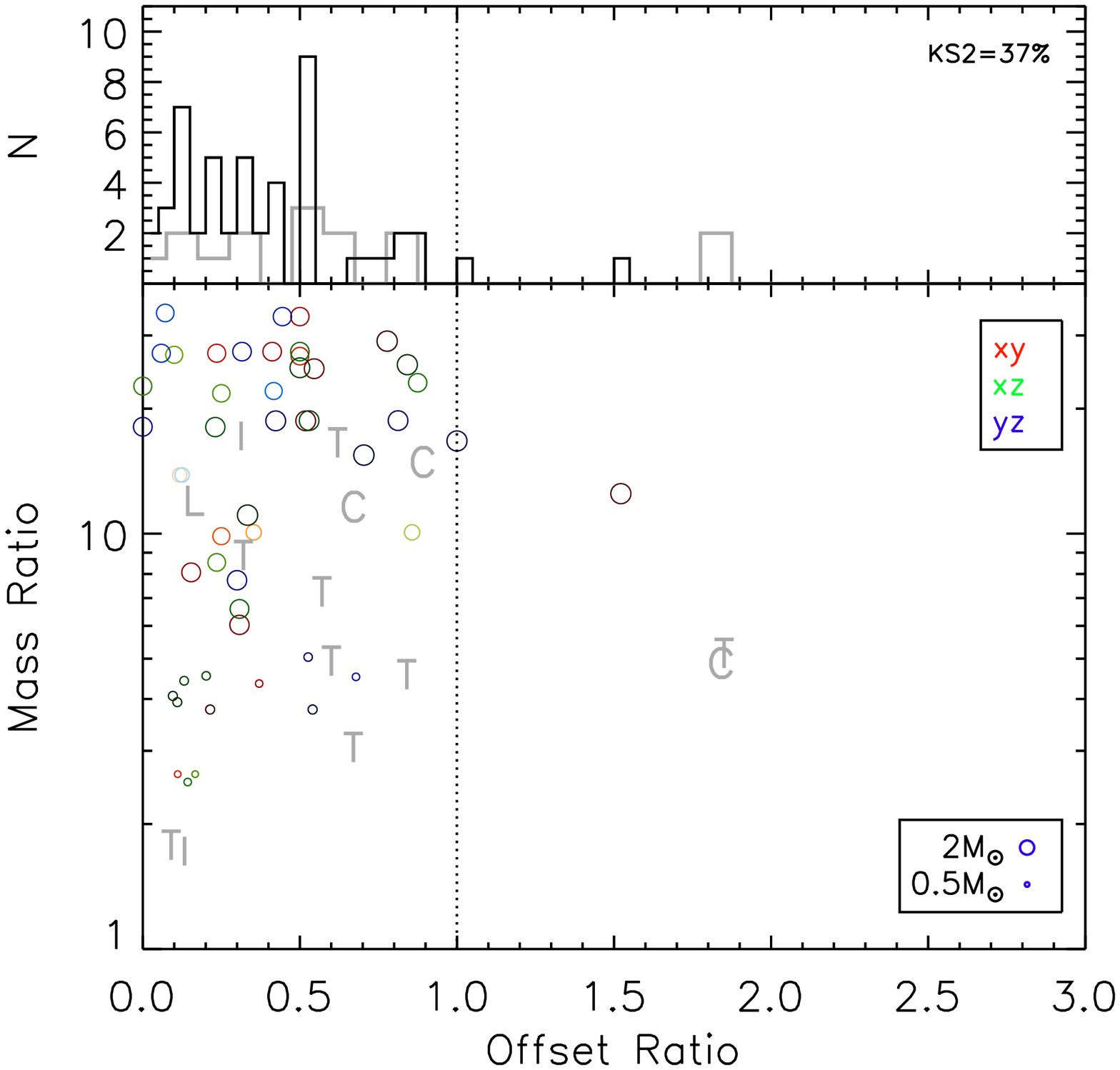} \\
\includegraphics[width=7.0cm]{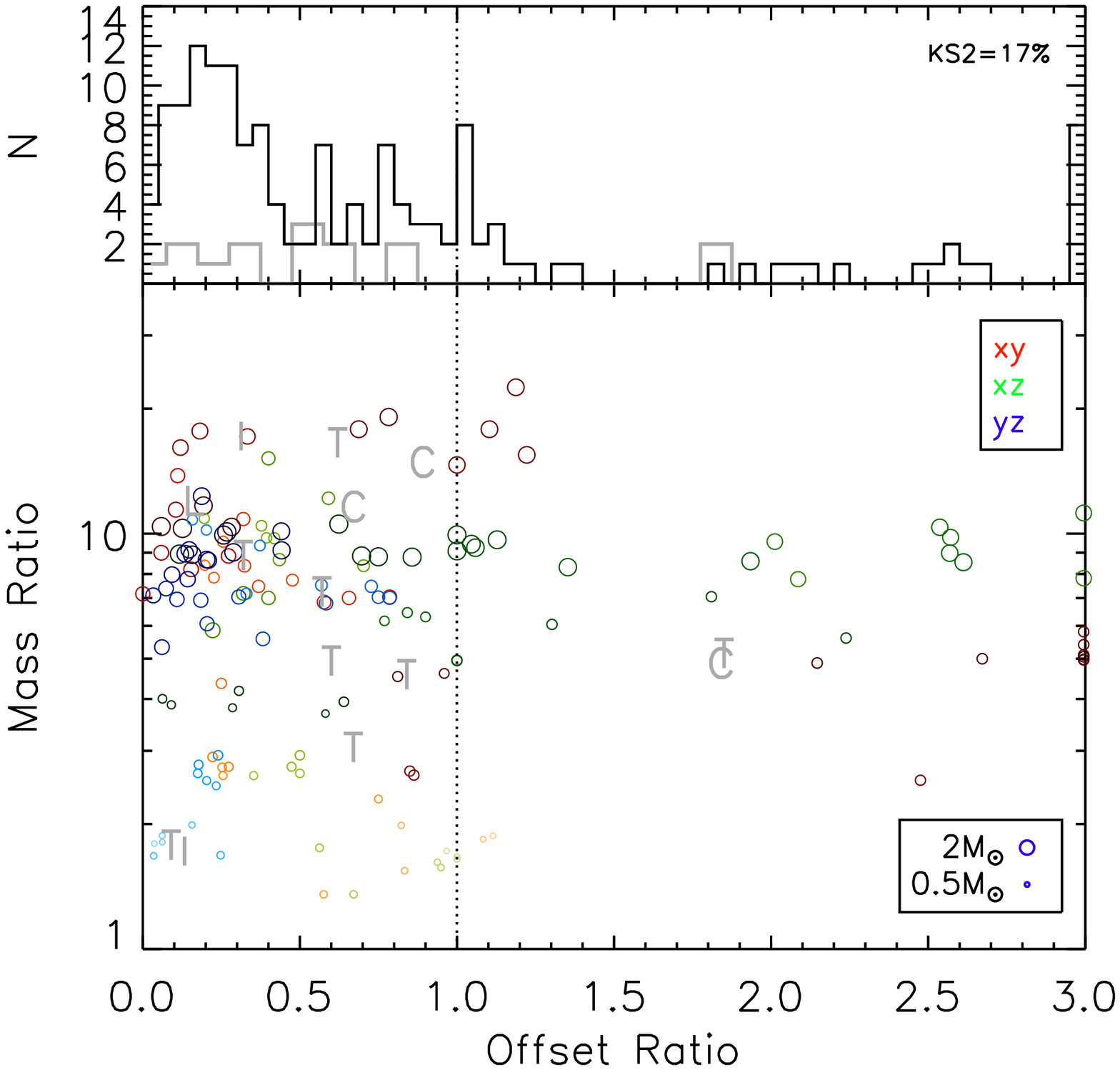} &
\includegraphics[width=7.0cm]{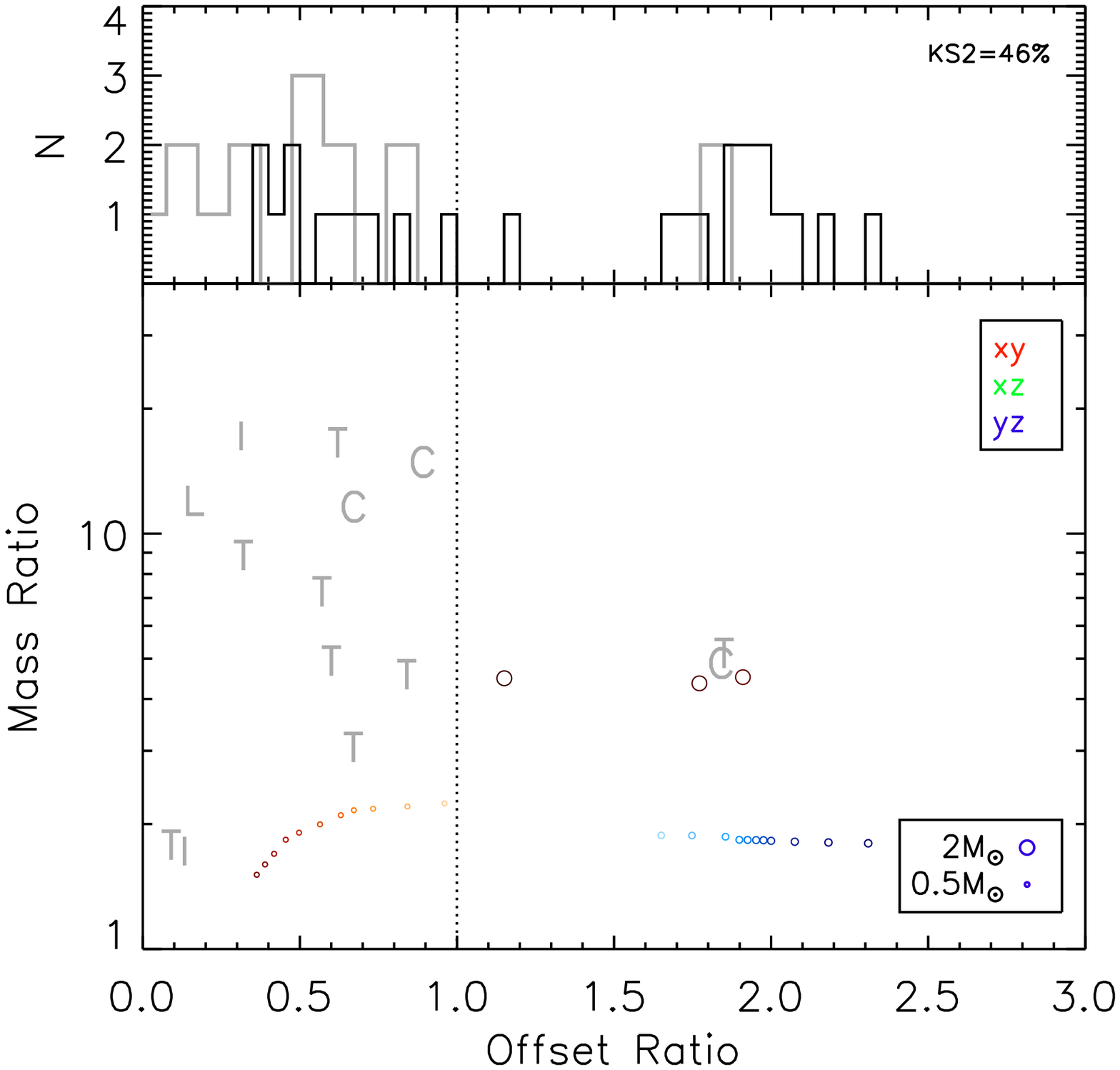} \\
\includegraphics[width=7.0cm]{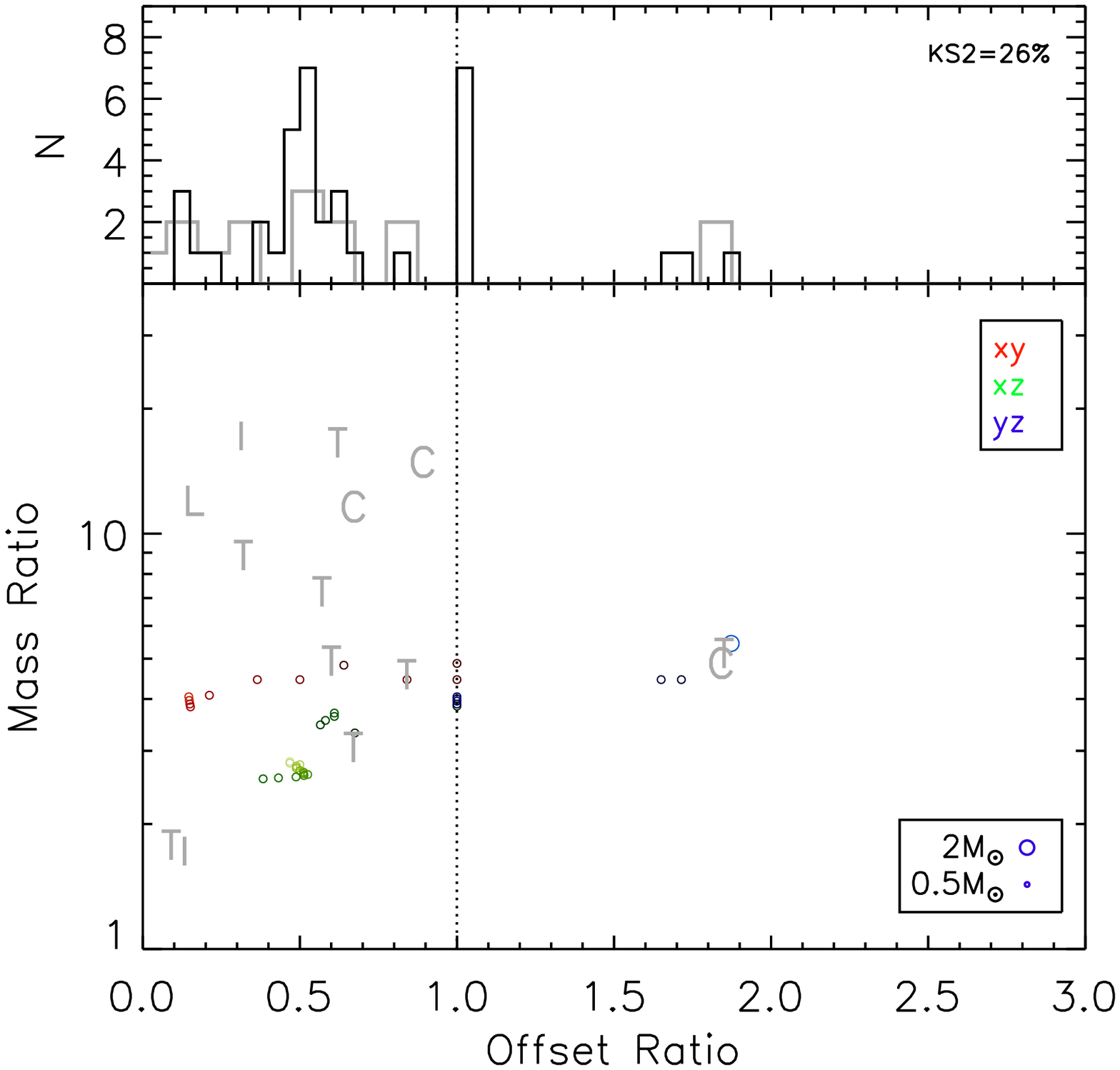} &
\includegraphics[width=7.0cm]{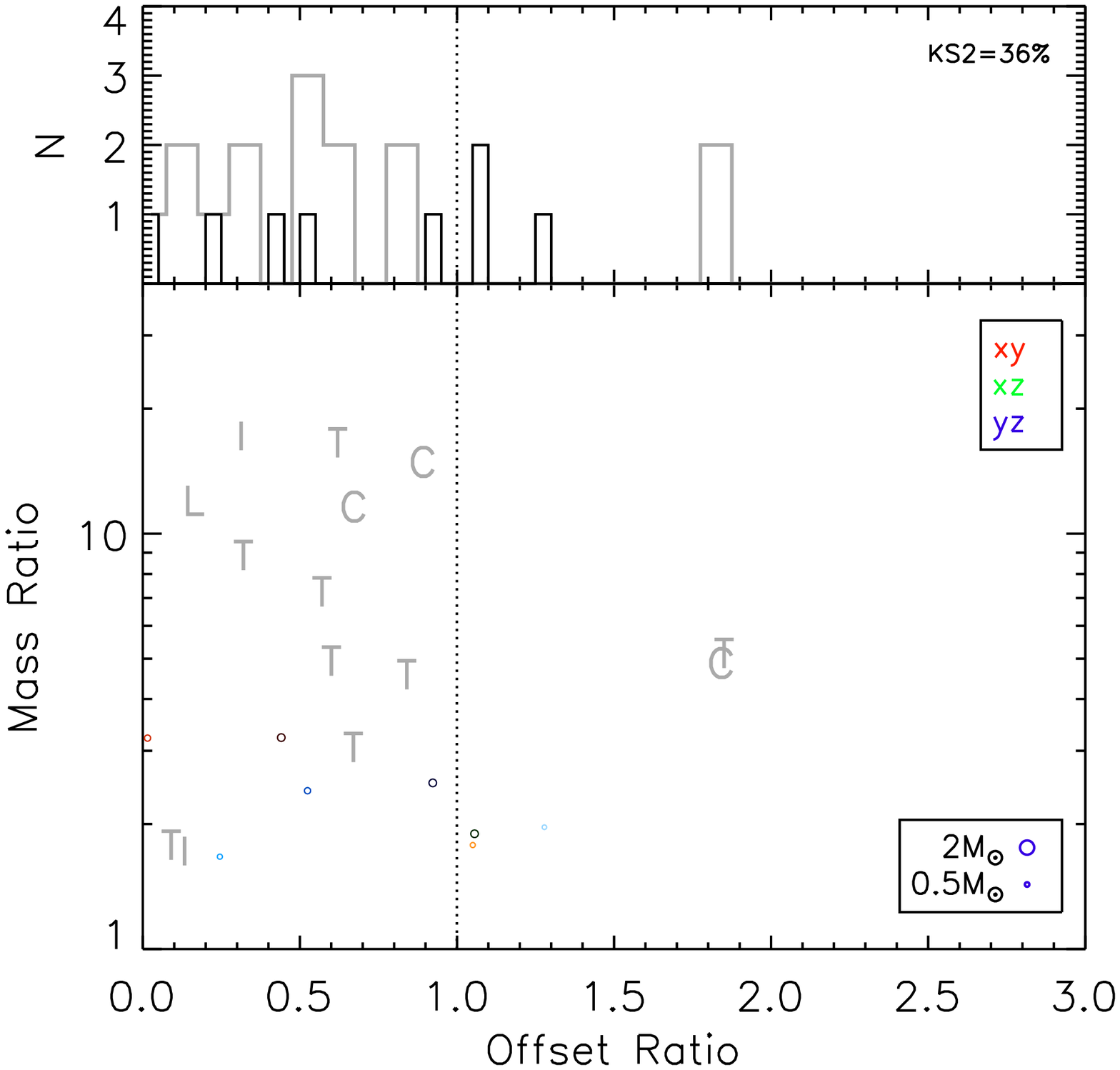} \\
\end{tabular}
\caption{The mass and offset ratios for clusters identified 
	with relaxed cluster definitions. Top
	row: Rk34 and Rm4, middle row: Rm6s and Rm9, bottom row: 
	Rt20 and Rrt with \Lcr values of 0.5, 0.15, 0.15, 0.5,
	0.5, and 0.15~pc respectively, and a minimum of 6 members
	required.  Clusters with offset
	ratios above 3 are plotted with a value of 3.
	For all simulations except Rm4 and Rm6s, {\it all} time slices
	in the simulation are included to increase the total number
	of data points.
	See Figure~\ref{fig_Rm6_ratios} for the
	plotting conventions used.}
\label{fig_all_ratios_extra}
\end{figure*}

Figures~\ref{fig_Rm6_ratios_extra} and \ref{fig_all_ratios_extra}
show the resulting mass and offset ratios.  In order 
to increase the amount of data available to plot,
the 0.01~Myr time sampling has only been applied to Rm4, Rm6, and
Rm6s; for the other simulations, all time steps are displayed.
For Rm6 and Rm6s,
these figures can be compared directly to their counterparts
in Figures~\ref{fig_Rm6_ratios} and \ref{fig_Rm6seed_ratios}.
Despite the substantial change in \Lcrns, as well as allowing 
clusters to be up to half as small, it is clear that the
results are qualitatively similar.  For both cluster definitions,
the majority of offset ratios are less than 1, and the distributions
are consistent with the observed clusters.  The mass ratios 
extend to smaller values with the maximally inclusive cluster
definition, since the smallest, sparsest, `clusters' now added
to the sample tend to lack the more massive stars present in the
original cluster sample.

After Rm6 and Rm6s, Rm4 produces the most clusters, and
clearly follows a similar trend to the former.  The remaining
simulations, despite the extremely relaxed cluster criteria adopted,
still form very few clusters.  Most of these 
have offset ratios of less than one.  The one
exception is Rm9, which appears equally split between
offset ratios below and above one.  Here,
only two clusters were identified (one each in the {\it xy} and {\it yz}
projections).  Both groupings
are small, and sufficiently sparse that they would likely not
have been visually selected as clusters.  Most of the identified
members have similarly small masses, which is a regime in which
the offset ratios could be expected to be random.  In the $xy$
projection (red points), at early times, a more massive star is 
separated by just under 0.5~pc from the cluster outskirts, leading
to the larger offset and mass ratios, both of which drop significantly
when that star drifts slightly further away.

\section{Cluster Centre Definition}\label{app_meancentre}
The ratio of the offset of the most massive cluster member from the
centre to the median cluster member offset, the {\it offset ratio},  
is influenced by one factor in addition to the cluster 
definition examined in Appendix~\ref{app_critlen},
the definition of the cluster's centre.  We demonstrate here
that this does not strongly influence on our results.

\begin{figure*}
\begin{tabular}{ccc}
\includegraphics[width=5.5cm]{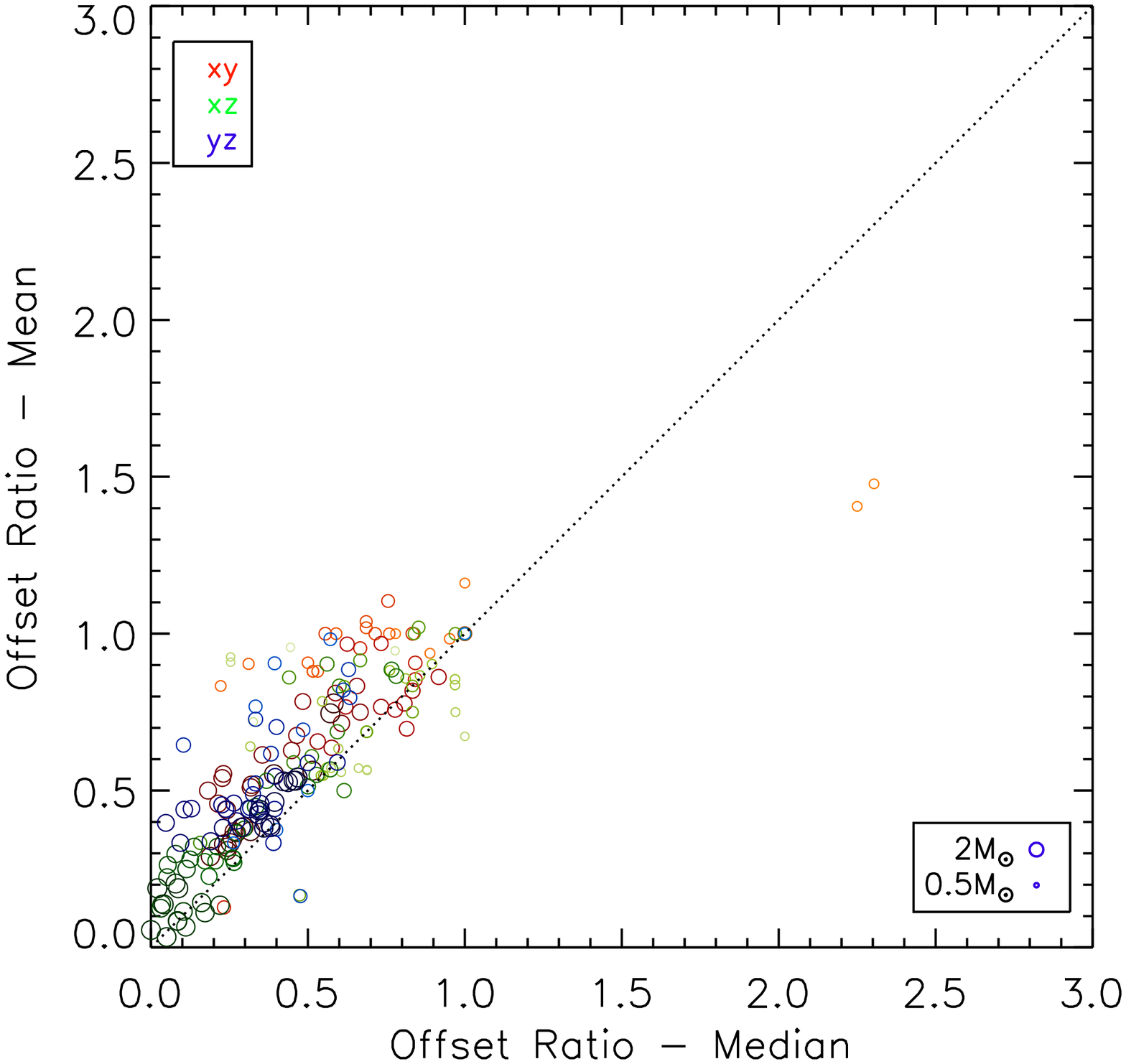} &
\includegraphics[width=5.5cm]{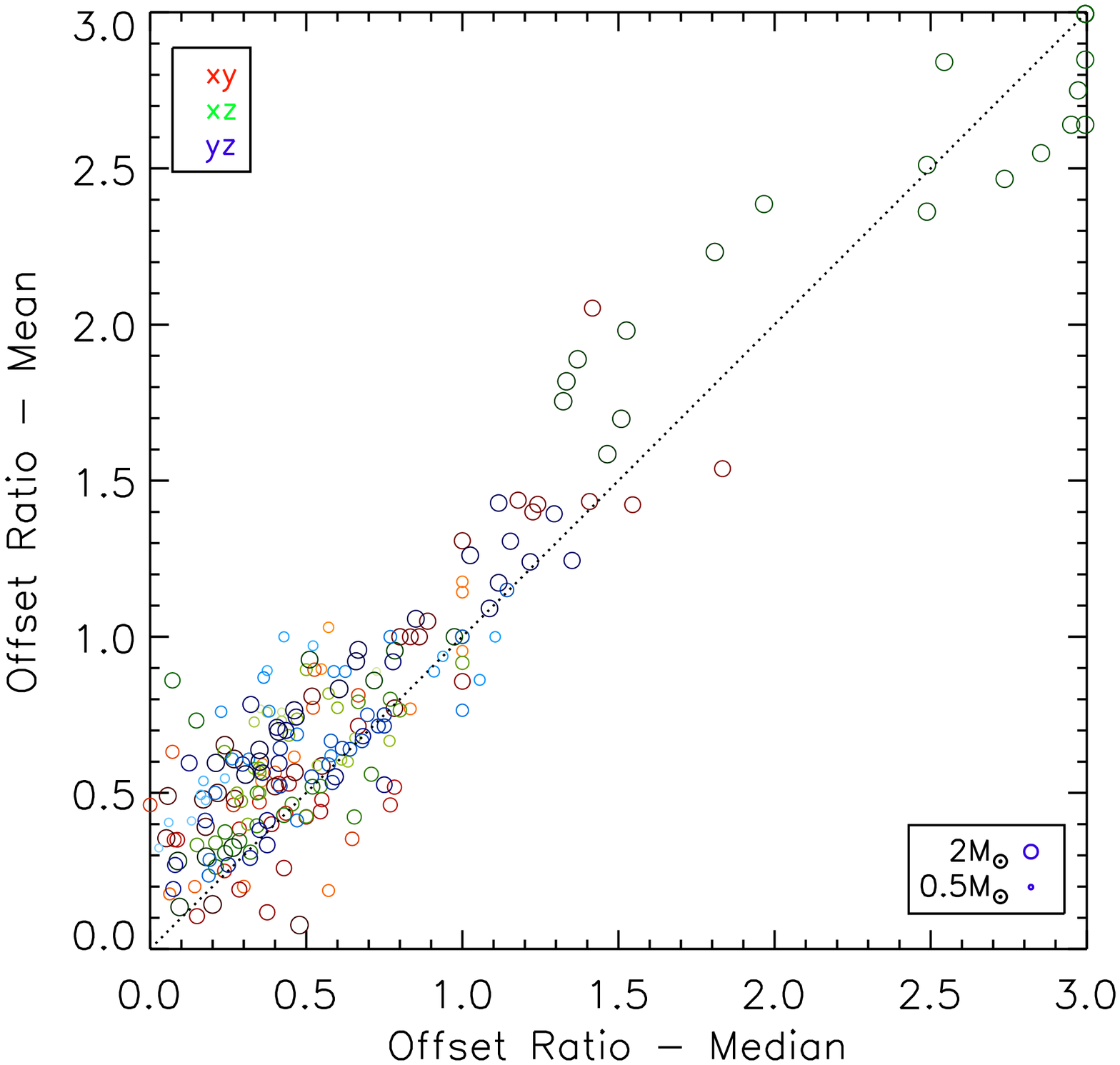} &
\includegraphics[width=5.5cm]{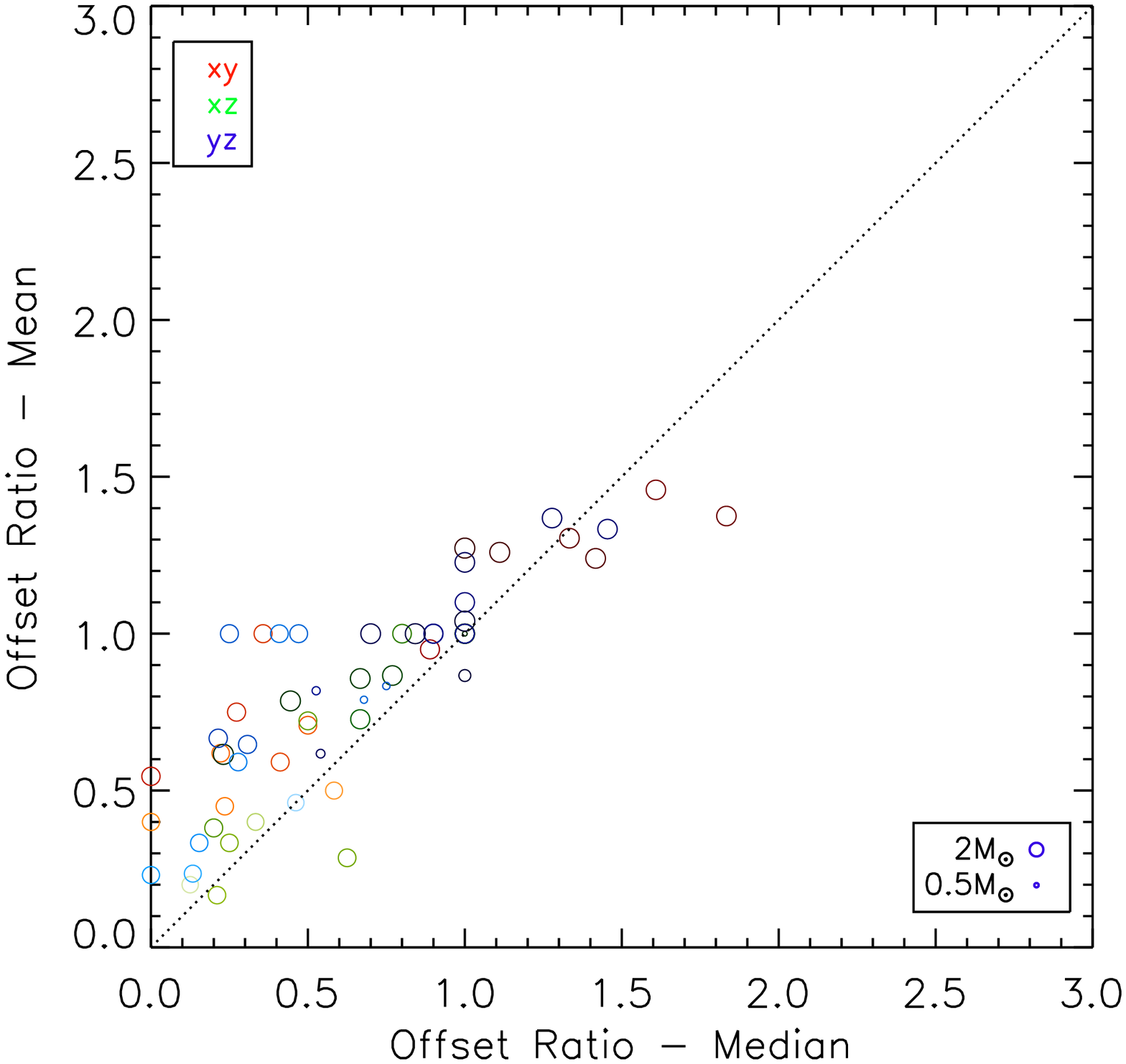} \\
\end{tabular}
\caption{A comparison of the offset ratios derived assuming the cluster centre 
	is the median or mean cluster member position.  From left to right: 
	Rm6, Rm6s, and Rm4.  The dashed line denotes a 1-1 relationship.
	As in Figure~\ref{fig_Rm6_ratios}, the colours denote the 
	projection in which the simulations are being viewed, the
	shading scales with the cluster's evolution in time, and
	the size of the symbols scales with the mass of the most massive
	cluster member.}
\label{fig_offset_comp}
\end{figure*}

Following \citetalias{kirk11},
in our main analysis, we adopt the median cluster position as the
centre, as the median is less
influenced than the mean by outliers in small-number samples. 
Here, we show the effect of instead adopting the mean position for
the centre.  Figure~\ref{fig_offset_comp} shows a comparison
of the offset ratios measured using the median- and mean-determined
centres, using \Lcr from Table~\ref{tab_critlens} and $N_{min}$ of 11
for Rm6 and Rm6s, and 6 for Rm4.
Figure~\ref{fig_offset_comp} shows significant scatter between the
two offset ratios, but no tendency for
the mean-determined offset ratio to be systematically higher
than the median-determined offset ratio (points lying
preferentially below the dashed line in Figure~\ref{fig_offset_comp} ).  
Instead, any bias appears to be in the opposite direction, i.e., 
with a mean-defined cluster centre, we would have
found preferentially lower offset ratios.  The small
excess of offset ratios of 1 is from 
clusters with an odd number of members, whose most massive
member has an offset ratio at precisely the middle of the range
of offset ratios.

Finally, we note that using the centre of mass of the cluster to define its
centre, there would more strongly tend to give small offset
ratios, since the central position would be biassed toward the
location of the most massive cluster member.
Our finding that the most massive
cluster member tends to form (and remain) in the cluster centre
is therefore robust to any reasonable variation imposed on the definition
of a cluster (previous section) and its centre (this section).

\section*{Acknowledgements}
The authors thank the referee, Ian Bonnell, as well as 
Thomas Maschberger and Cathie Clarke for helpful suggestions.
The authors thank Phil Myers for interesting discussions which
inspired several of the figures presented here.
The authors also thank Jonathan Tan, Sourav Chatterjee,
and Cara Battersby for helpful discussions.  This research has
been supported by the Smithsonian Scholarly Studies Program (HK),
the Smithsonian Astrophysical Observatory (HK, KR), 
the NSF through grant AST-0901055 (SSRO), and the NSF REU and
DOD ASSURE programs under NSF grant no.~0754568 (KR). 
HK acknowledges support from the Banting Postdoctoral Fellowship,
administered by the Government of Canada. 
Support for this work was provided by NASA through Hubble 
Fellowship grant \#51311.01 awarded by the Space Telescope Science 
Institute, which is operated by the Association of Universities for 
Research in Astronomy, INC., for NASA, under contract NAS 5-26555 (SSRO). 
The simulations were performed on the XSEDE Trestles resource (SSRO).

\bibliography{clusterbib}{}
\bibliographystyle{mn2e}

\end{document}